\documentclass[fleqn,10pt]{wlscirep}
\usepackage[utf8]{inputenc}
\usepackage[T1]{fontenc}
\usepackage{float}
\usepackage{multicol}
\usepackage{tikz}
\usepackage{xcolor}
\usetikzlibrary{arrows.meta, positioning, fit, calc}
\usepackage{multirow}
\usepackage{booktabs}
\usepackage{comment}
\usepackage{url}
\usepackage{subcaption}
\usepackage{hyperref}

\newcommand{\added}[1]{\textcolor{blue}{#1}}

\title{State-space representation of early-life child mortality
dynamics: linking event history analysis and orbit-based models}

\author[1]{Wendy Lefotlha}
\author[1]{David Sherwell}
\author[2,3,*]{Maria Vivien Visaya}

\affil[1]{School of Computer Science and Applied Mathematics,
University of the Witwatersrand, Braamfontein, 2000, South Africa}
\affil[2]{Department of Mathematics and Applied Mathematics,
University of Johannesburg, Johannesburg, South Africa}
\affil[3]{National Institute for Theoretical and Computational
Sciences (NITheCS), South Africa}
\affil[*]{mvvisaya@uj.ac.za}

\begin{abstract}Understanding early-life child mortality requires methods 
that capture both statistically significant risk factors 
and the structure of individual life trajectories. Event 
History Analysis (EHA) estimates associations between 
covariates and mortality risk but may obscure rare or 
high-dimensional configurations of risk. Orbit Theory (OT) 
represents individuals as exact trajectories in a 
combinatorial state space, preserving full multivariate 
structure. We analyse longitudinal data across 12 variables 
for 31,081 children from the Agincourt Health and 
Demographic Surveillance System (South Africa, 1998--2008) 
using both methods. We show three things EHA alone cannot. 
First, maternal migration is structurally pivotal in the 
transition dynamics leading to child death despite being 
statistically non-significant in EHA. Second, maternal 
refugee status, while statistically significant, generates 
no trajectory dynamics and is correctly excluded from any 
structural reduction. Third, a three-variable subsystem 
(child status, mother status, mother migration) recovers 
EHA's principal findings within a reduced state space of 
at most 48 states while exposing small high-risk clusters 
invisible to regression-based analysis. Maternal death is 
the dominant EHA predictor; maternal migration organises 
87\% of the observable sequential transitions into 
child-death states. These are not contradictory: maternal 
death rarely appears as a preceding state because its role 
is terminal rather than sequential. Early-life mortality 
can be read at two levels simultaneously -- statistically, 
through effect estimates, and structurally, through observed 
transitions. 

\textbf{Keywords:} under-five mortality, event history analysis,
orbit theory, longitudinal data, state space dynamics
\end{abstract}

\begin{document}
\flushbottom
\maketitle

\section{Introduction}

Under-five mortality remains a central concern in population health
and demographic research, despite substantial global declines over
recent decades~\cite{You2015,Liu2016,UNIGME2024}. Recent estimates
indicate that approximately 4.8 million children died before the age
of five in 2023, with the majority of deaths occurring in low- and
middle-income countries~\cite{SDG2023}. Persistent inequalities in
under-five mortality across sub-Saharan Africa are driven by
socio-economic disparities and household-level
vulnerabilities~\cite{Bocquier2011,Clark2013,Houle2021}. Mortality
is strongly concentrated in the earliest years of life, reflecting a
combination of biological vulnerability and exposure to adverse
household and environmental conditions~\cite{You2015,UNIGME2024}.

The Agincourt Health and Demographic Surveillance System (HDSS) in
rural South Africa provides a detailed longitudinal platform for
examining these dynamics~\cite{kahn2012}. The site has been widely
used to study demographic processes including mortality, migration,
and household structure~\cite{kahn2006,Clark2007}. Previous studies
have shown that child survival is strongly influenced by maternal
survival, household composition, and broader socio-demographic
factors~\cite{Bocquier2011,Clark2013,Houle2013,Houle2015,Houle2021}.
 Understanding under-five mortality requires modelling processes that
are  dynamic, multivariate, and structured over time.
Children do not experience risk in isolation; rather, mortality
emerges from evolving configurations of maternal survival, migration,
household composition, and sibling events. 
Each child traces a trajectory through a binary state space defined by these configurations.   
 Standard approaches such as
discrete-time Event History Analysis (EHA) model this process
through hazard functions, yielding statistically interpretable
estimates of covariate effects on the probability of death. This
representation is  arguably reductive: it compresses the joint
evolution of variables into states and events such as death using a parametric form, inadvertently masking rare or
structurally distinct configurations of risk.

OT is not an alternative inferential method to EHA. EHA provides
statistical inference  -  effect estimates, significance tests,
and adjustment for confounding and unobserved
heterogeneity~\cite{Singer2003,BoxSteffensmeier2004,yamaguchi1991,
Allison1984}  -  but operates through aggregation and may obscure
the detailed structure of individual trajectories. Sequence-based
approaches in the social sciences have long flagged this limitation,
emphasising the importance of preserving trajectory structure rather
than aggregating over states~\cite{Abbott1995,Billari2001,Aassve2007}.
OT addresses it directly: it represents each individual as an exact
trajectory in a discrete state space, preserving the complete
configuration of variables and their temporal evolution, and making
transition mechanisms directly observable~\cite{Visaya2014,Visaya2015}.
Unlike sequence analysis, OT provides an explicit state-space
representation in which transition mechanisms can be analysed as
dynamical objects~\cite{Massonis2023}. 
  This study constructs a reduced orbit system that links EHA 
and OT within a unified statistical state-space framework. 

Using longitudinal data across 12 variables for 31,081 children from
the Agincourt HDSS (1998--2008), we show three things. First, that maternal migration is structurally pivotal in the
transition dynamics leading to child death despite being statistically
non-significant in EHA. Second, that maternal refugee status, while
statistically significant, generates no trajectory dynamics and is
correctly excluded from any structural reduction; and third, that a
three-variable subsystem $(CS, MS, MM)$ recovers EHA's principal
findings within a reduced state space of 48 states while additionally
exposing  
 clusters invisible to regression-based
analysis.  
  This reduced three-variable orbit system is the central contribution
of OT;  it facilitates comparison with EHA. It is treated as a methodological object in its own
right: a structured projection of the full multivariate binary orbit
space onto a lower-dimensional subsystem that preserves the principal
mortality dynamics identified by EHA, while simultaneously exposing
trajectory configurations and small high-risk clusters that are
invisible to regression-based analysis. 
  We identify this subsystem through a projection  $\Phi: \{0,1\}^n \to \{0,1\}^3$, mapping each child's full  binary state vector onto the three variables $(CS, MS, MM)$  at each age.
 The mapping $\Phi$ is constructed after inspection of variable
persistence and association with adverse outcomes.  It is therefore
data-informed but not model-dependent.   

From a methodological perspective, the relationship between EHA and
OT can be understood as a distinction between probabilistic and
state-space representations of the same underlying process. EHA
models the transition from survival to death through a conditional
probability model, estimating  $\mathbb{P}(CS_t = 0 \mid x_t)$,  
where $x_t$ denotes the covariate configuration at time $t$. OT
represents the evolution of $x_t$ itself as a trajectory in a
discrete state space, making transitions between configurations
 observable. The two approaches operate on the same
underlying system but at different levels of abstraction: EHA
provides statistical inference on transition likelihoods, while OT
provides a structural description of how individuals move between
mortality-related states over time. Recent state-space approaches to
longitudinal survival analysis have similarly highlighted the
importance of representing temporal evolution directly within the
underlying system dynamics rather than solely through marginal hazard
estimates~\cite{Cauchi2024}. 
The reduced orbit system introduced here is not a dimensionality
reduction technique in the sense of principal component analysis or
manifold learning. Rather, it is a trajectory-based representation of
mortality dynamics that preserves the significant transition mechanisms
governing child death while collapsing configurations that do not
contribute to mortality behaviour. The goal is structural fidelity to
the transition dynamics, not variance preservation or reconstruction
accuracy.

  As a contribution of this study, first, it applies
both EHA and OT to the same demographic dataset, isolating differences
that arise from the choice of representation rather than from the
data. EHA operates through parametric modelling of the hazard function
and yields statistically interpretable effect sizes; OT represents
trajectories as deterministic paths in a combinatorial state space,
preserving full variable configurations. 
Second, it introduces a complementary statistical--reduced orbit
system representation in which EHA and OT operate at different but
compatible levels of abstraction. EHA provides statistically validated
effect estimates and controls for confounding and unobserved
heterogeneity, identifying a minimal set of relevant covariates. OT,
applied to the reduced set, exposes the structural organisation of
trajectories, including disjoint clustering of significant and rare but persistent configurations not recovered within regression-based models.

\section{Data and Methods}

This study involves secondary analysis of existing de-identified 
demographic surveillance data and does not involve experiments on 
human subjects. All methods were carried out in accordance with 
relevant guidelines and regulations, and all data collection 
protocols were approved by the Human Research Ethics Committee 
(Medical) of the University of the Witwatersrand, Johannesburg, 
as described in Kahn(2012)\cite{kahn2012}. Informed consent was obtained 
from all participating households by the Agincourt HDSS at the 
point of enumeration.
 The data,  drawn from the Agincourt
HDSS~\cite{kahn2012, Clark2007,kahn2006}, is a longitudinal
population-based surveillance system located in rural South Africa.
The study population comprises approximately 70,000 individuals across
21 villages, with continuous demographic monitoring since 1992. For
this analysis, we consider 31,081 children observed between 1998 and
2008. 
 Children are
followed from entry into the study until death, out-migration, or
reaching age five. The outcome variable is a binary indicator of child
death within each discrete time interval. Time is sequenced in years, 
by analysis ages $a \in \{0,1,2,3,4\}$. Covariates include maternal
death, maternal migration, maternal refugee status, maternal age at
birth, child gender, and sibling-related variables constructed from
birth and death records. These variables are coded in binary form to
enable comparability between EHA and OT.

\subsection{Data representation in EHA and OT}
\label{sec:datastructuring}

For EHA, the data are structured in person-period format, where each
child contributes one record per year of observation until death,
censoring, or exit from the study. The response variable is binary,
indicating whether death occurs in a given interval, and covariates
may be time-varying or fixed. This representation supports estimation
of discrete-time hazard models. For OT, the same data are represented
as sequences of binary state vectors. For each child $k$, the
longitudinal record is encoded as
$$
x_t^{(k)} \in \{0,1\}^n, \quad t = 0,1,2,3,4,
$$
where $n$ is the number of variables. In the full analysis, $n = 12$,
corresponding to the twelve binary variables listed in
Table~\ref{tbl:vars}. These sequences are then reordered according to
variable-specific frequencies of change to construct orbit states
$(\mathbf{x}_t, \mathbf{y}_t)$ in the state space $S_n = X_n \times
Y_n$ (see Section~\ref{sec:orbit_theory} for full details). While EHA
represents the data through conditional probabilities of event
occurrence, OT represents the same data as trajectories through a
structured state space. The reduced orbit system introduced in
Section~\ref{sec:reduced} is constructed from this representation. 


All variables are treated as binary in OT, where $1$ denotes a
favourable (fit) value and $0$  unfavourable (unfit). In
contrast, EHA retains categorical structure where appropriate (e.g.,
mother age groups) and incorporates these as dummy variables within
the regression model while OT codes each category as zero or one. The resulting dataset is a multivariate binary
longitudinal dataset (BMLD), observed at discrete time points
corresponding to child age in years. Each child contributes a sequence
of binary vectors across at most five observation periods.  Table~\ref{tab:otvars} lists the variable encodings.

Two terms used throughout this paper require precise definition
within the OT framework. A \textit{configuration} refers to a specific
realised value of the binary state vector $x_t \in \{0,1\}^n$ at a
given time point  -  the complete joint assignment of all $n$ variables
for a given child at age $t$. Two children share a configuration at
age $t$ if and only if their binary state vectors are identical across
all $n$ variables simultaneously; it is therefore a joint property of
the full variable set, not a marginal property of any single variable.
A \textit{high-risk} state or cluster is defined empirically as a
configuration or set of configurations from which at least one
observed trajectory transitions into a child-death absorbing state
($CS_t = 0$) within the observation window, weighted by the count of
children occupying that configuration. High-density states among
these are of direct public health relevance; low-density (singleton
or near-singleton) states represent rare but structurally distinct
high-risk subgroups. Neither definition relies on a parametric model
or a significance threshold: both are read directly from the observed
transition structure of the data.

\begin{table}
\centering
\caption{Binary encoding of variables for EHA and OT. A 1 denotes a favourable state for child survival while 0 denotes an unfavourable state.}
\label{tab:otvars}
\begin{tabular}{lll}
\toprule
\textbf{Variable} & \textbf{EHA value} & \textbf{OT encoding} \\
\midrule
Mother age group & $15$--$24$ & 0 Unfavourable \\
& $25$--$59$ & $1$ Favourable \\
Mother refugee status (RS) & Mozambican / South African
  & Mozambican ($0$) / SA ($1$) \\
Mother migration (MM) & Out / Same residence
  & Out MO ($0$) / In MI ($1$) \\
Mother status (MS) & Dead / Alive
  & Dead MD ($0$) / Alive MA ($1$) \\
Child age & $0,1,2,3,4$ & Discrete time index \\
Child gender & Male / Female & Male ($0$) / Female ($1$) \\
Succeeding sibling birth (SSB) & Yes / No & Yes ($1$) / No ($0$) \\
Succeeding sibling death (SSD) & Yes / No & Death ($0$) / No ($1$) \\
Preceding sibling death (PSD) & Yes / No & Death ($0$) / No ($1$) \\
\bottomrule
\end{tabular}
\end{table}


\subsection{Event History Analysis}
\label{sec:eha}

EHA is implemented using discrete-time logistic
regression~\cite{Singer2003,BoxSteffensmeier2004}. The outcome
variable is child death. The discrete-time hazard model is specified as
$$
\text{logit}\,\mathbb{P}(CS_{it}=0) = \alpha_a + \beta^\top X_{it}
+ u_i,
$$
where $\alpha_a$ denotes the baseline hazard for analysis age $a$,
$X_{it}$ is the vector of covariates for child $i$ at time $t$,
$\beta$ is the corresponding coefficient vector, and $u_i$ is a
child-specific random effect capturing unobserved heterogeneity. The
frailty term $u_i$ is assumed to follow a normal distribution with
mean zero and variance $\sigma_u^2$. Frailty models of this type are
standard for discrete-time survival data with repeated observations
per individual~\cite{rabe2008multilevel,yamaguchi1991}, and
their use here is motivated by the expectation that unmeasured
child-level factors (e.g., birthweight, nutrition, access to care)
will induce residual correlation across person-periods. All
models were fitted in Stata using \texttt{melogit} with
\texttt{vce(bootstrap, reps(500))}, providing bootstrap standard
errors robust to influential observations. The overall model
significance was assessed using a Wald $\chi^2$ test with degrees of
freedom equal to the number of covariates in the respective model.

The explanatory variables are biological mother death, biological
mother temporary out-migration, biological mother refugee status,
biological mother age group at the child's birth, child gender, child
age, and sibling-related variables. The sibling variables distinguish
preceding sibling death, succeeding sibling death, and succeeding
sibling birth, each examined in both short-term (\emph{shock}) and
longer-term (\emph{trailing}) forms under a one-year lag structure.
Child age is represented by five annual categories, indexed as
analysis age $a \in \{0,1,2,3,4\}$, where $a=0$ denotes the first
year of life and serves as the reference category. Mother's age at
birth is grouped into the categories $15$--$24$, $25$--$29$,
$30$--$34$, $35$--$39$, and $40$--$59$, with $15$--$24$ as the
reference group.

\subsection{Orbit Theory}
\label{sec:orbit_theory}

Orbit Theory represents each individual as a trajectory in a discrete
state space~\cite{Visaya2014}
$$
S_n = \mathcal{X}_n \times \mathcal{Y}_n,
$$
where $\mathcal{X}_n = \{0,1\}^n$ encodes the binary configuration
and $\mathcal{Y}_n$ encodes the reordering of $n$ variables according
to persistence.  The theoretical cardinality of $S_n$ is $|S_n| = 2^n \cdot n!$, 
reflecting all possible combinations of binary configurations 
and variable orderings. In practice, observed trajectories 
occupy a sparse subset of this space: for $n=12$, 
$|S_{12}| = 2^{12} \cdot 12! \approx 2 \times 10^{10}$, 
yet only 329 states are observed in the data.  Each child $k$ is represented at age $t$ as
an ordered pair
$$
(x_t^{(k)},\, y_t^{(k)}), \quad t = 0,1,2,3,4,
$$
in $S_n$. The construction proceeds as follows. Each variable is
binary-encoded: 1 if the state is favourable for child survival, 0
otherwise. Variables that do not change over the observation period
for a given child (e.g., gender, maternal refugee status) are placed
first in the ordering. Changing variables (e.g., maternal migration,
maternal death) are then arranged in increasing order of their
population-level frequency of change, so that the most slowly-changing
variables appear to the left (most significant digits) and the most
rapidly-changing variables appear to the right (least significant
digits). This ordering defines the $x$-coordinate, referred to as
the \textit{fitness axis}. For $n$  variables, a
digit string  $0^n$ of length $n$ denotes the most unfit state and $1^n$ the
completely fit state. The $y$-coordinate is defined by the specific
permutation of variable names used; there are $n!$ possible such
permutations, and this axis is referred to as the
\textit{significance axis}. A single change in a child's variable values from age $t$ to age $t+1$ drives its trajectory in $S_n$: a changing variable is pushed to the rightmost (least significant) position, progressively exposing the most persistent variables.  When two or more variables change simultaneously in a single 
transition, their relative significance positions are 
preserved in the swap. The full formal definition of the 
swap rule, including the simultaneous-change case, is given 
in~\cite{Visaya2014}.  Maternal death, for example, persists across time steps rather than cycling, and so moves leftward in the ordering over the observation window. 

Orbit Theory can be understood as a sequence-analytic approach
that, rather than summarising trajectories by optimal-matching distance
or state frequencies, represents each individual as an exact path in a
fully enumerated discrete state space, making transition mechanisms
directly
observable~\cite{Abbott1995,Billari2001,Aassve2007}.  The full analysis uses $n = 12$ variables, listed in
Table~\ref{tbl:vars}. 
  Trajectories are observed
directly as paths through the state space, making it possible to
identify clusters, state transitions, and configurations in which
orbits remain for extended periods of time.

\section{Results}

\subsection{EHA Results}
\subsubsection{Descriptive mortality pattern}

\begin{figure}
\centering
\includegraphics[width=0.5\textwidth]{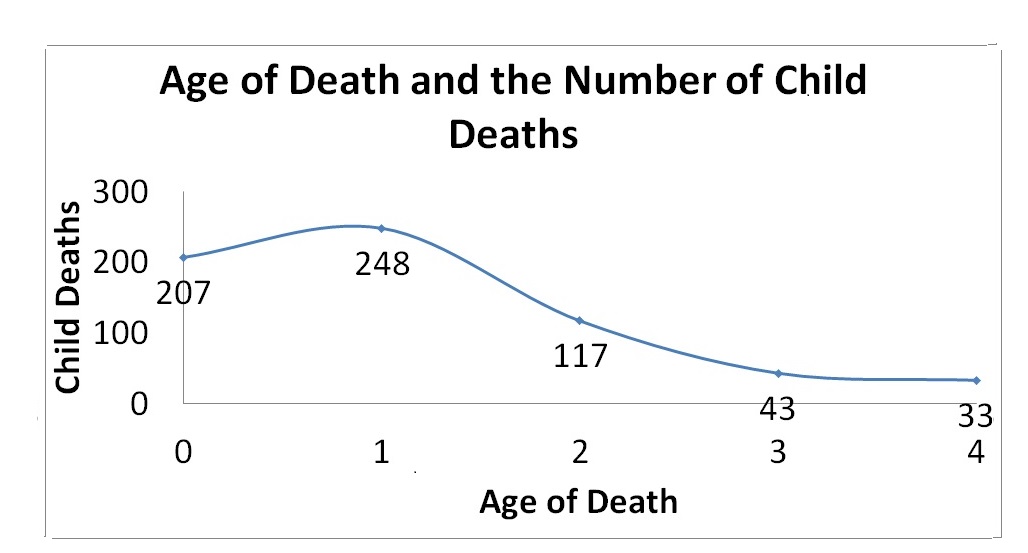}
\caption{Age-specific mortality pattern. }
\label{fig:death-age}
\end{figure}

Table~\ref{tab:ageanddeath} and Figure~\ref{fig:death-age} report the
age distribution of deaths across the full study period. Across the
full study period, 648 child deaths are recorded. The largest number
of deaths occurs at analysis age 1 ($n = 248$), followed by 207
deaths at analysis age 0. Thereafter, the number declines steadily
with age. This non-monotonic profile  -  mortality rising from age 0
to age 1 before declining sharply  -  reflects annual aggregation of
the data rather than the neonatal-period peak commonly reported in
monthly-scale analyses. Mortality concentrated in early childhood is
consistent with global patterns~\cite{You2015,UNIGME2024}. The age-hazard correspondence between the EHA frailty model and
the $S_3$ transition counts (Table~\ref{tab:age_correspondence})
confirms that both representations agree on the concentration of
mortality risk in early childhood. 
%
 This
concentration corresponds to the high density of early-time
transitions observed in the orbit representation
(Section~\ref{sec:orbit_theory}), and is consistent with the positive
frailty-model coefficient at age 1 followed by strongly negative
coefficients at ages 2--4 in Table~\ref{tab:f}.

\begin{table}
\centering
\caption{Number of children alive and dead by analysis age.}
\label{tab:ageanddeath}
\begin{tabular}{cccc}
\toprule
Analysis age & Alive & Died & Total \\
\midrule
0 & 15\,725 & 207 & 15\,932 \\
1 & 16\,096 & 248 & 16\,344 \\
2 & 15\,798 & 117 & 15\,915 \\
3 & 15\,383 & 43  & 15\,426 \\
4 & 14\,977 & 33  & 15\,010 \\
\midrule
Total & 77\,979 & 648 & 78\,627 \\
\bottomrule
\end{tabular}
\end{table}

\subsubsection{Sibling-exposure structure}

Before fitting the regression models, it is useful to note the
exposure structure of the sibling variables. 
 Table~\ref{tab:prsnyrs} shows that succeeding sibling birth
contributes far more person-years than either preceding or succeeding
sibling death. This
matters for interpretation, because sparse exposure combined with
lagging can generate unstable or directionally implausible
coefficients. This asymmetry helps explain why sibling birth remains
the only sibling-related variable with a plausible positive
association in later models, while sibling death coefficients become
unstable and difficult to interpret substantively.
\begin{table}[H]
\centering
\caption{Person-years for sibling death and birth exposures.}
\label{tab:prsnyrs}
\begin{tabular}{lcccc}
\toprule
Variable & Trailing & Lag trailing & Shock & Lag shock \\
& person-years & person-years & person-years & person-years \\
\midrule
Succeeding birth  & 3\,152 & 2\,331 & 821 & 921 \\
Succeeding death  & 68     & 33     & 35  & 15  \\
Preceding death   & 92     & 61     & 31  & 23  \\
\bottomrule
\end{tabular}
\end{table}

\subsubsection{Shock and trailing models}

Two preliminary discrete-time logistic models are fitted without
frailty. The first assesses one-year lagged shock effects; the second
assesses one-year lagged trailing effects. These models evaluate
whether sibling-related variables add meaningful information beyond
the maternal and age-related covariates. 
Table \ref{tab:seffect}  
presents the
short-term (shock) model. 
The maternal and age-related covariates are stable across both.
Maternal death has the largest positive coefficient in each case,
consistent with previous studies~\cite{Clark2013,Houle2015,Houle2021},
and this dominance is reflected structurally in the orbit
representation, where transitions into child-death states concentrate
along trajectories associated with changes in maternal survival status
(Section~\ref{sec:reduced}). Refugee status remains positively
associated with child death, reflecting broader social
inequalities~\cite{Bocquier2011,Birhanie2025,Avelino2025}. 
Among the sibling variables, succeeding sibling birth shows a
positive coefficient in both models, with stronger evidence in the
trailing specification ($\exp(0.492)\approx 1.64$, $p=0.056$),
suggesting that the presence of a younger sibling is associated with
elevated mortality risk for the index child particularly beyond the
immediate year of birth. By contrast, the sibling death variables
produce very large negative coefficients. In the shock model, the
odds of child death associated with succeeding sibling death are
$\exp(-12.653)\approx 3.2\times10^{-6}$ relative to children without
a recorded succeeding sibling death; for preceding sibling death the
corresponding odds ratio is $\exp(-14.550)\approx 4.8\times10^{-7}$.
Both are statistically significant but substantively implausible.
Under the one-year lag structure, a child must survive through the
exposure window to be observed as at risk in the subsequent interval,
so the apparent protective effect is an artefact of survivorship bias
rather than evidence that sibling death lowers child mortality risk.
Sibling death variables are therefore not retained for substantive
interpretation in the final model. 
\begin{table}
\centering
\caption{Short-term (shock) effects model for child death. Succeeding sibling birth has no significant association with child death.}
\label{tab:seffect}
\small
\begin{tabular}{llccc}
\toprule
 & Variable & Coefficient & $p$-value & 95\% CI \\
\midrule
\multirow{4}{*}{Analysis age (base: 0)}
  & 1 &  $0.128$ & $0.198$ & $(-0.066,\ 0.323)$ \\
  & 2 & $-0.565$ & $<0.001$ & $(-0.797,\ -0.333)$ \\
  & 3 & $-1.588$ & $<0.001$ & $(-1.946,\ -1.230)$ \\
  & 4 & $-1.933$ & $<0.001$ & $(-2.338,\ -1.527)$ \\
\midrule
\multirow{3}{*}{Mother}
  & Migration  & $-0.045$ & $0.764$ & $(-0.345,\ 0.253)$ \\
  &  Status  & $3.257$ & $<0.001$
    & $(1.440,\ 5.069)$ \\
  & Refugee status & $0.333$ & $0.001$ & $(0.174,\ 0.492)$ \\
\midrule
\multirow{4}{*}{Mother age (base: 15--24)}
  & 25--29 & $0.293$ & $0.018$ & $(0.051,\ 0.535)$ \\
  & 30--34 & $0.297$ & $0.021$ & $(0.044,\ 0.550)$ \\
  & 35--39 & $0.170$ & $0.254$ & $(-0.122,\ 0.462)$ \\
  & 40--59 & $0.170$ & $0.228$ & $(-0.146,\ 0.466)$ \\
\midrule
& Gender & $-0.020$ & $0.796$ & $(-0.175,\ 0.134)$ \\
\midrule
\multirow{2}{*}{Succeeding sibling}
  & Birth & $0.353$ & $0.294$ & $(-0.306,\ 1.014)$ \\
  & Death & $-12.653$ & $<0.001$
    & $(-20.937,\ -4.336)$ \\
\midrule
Preceding sibling & Death & $-14.550$ & $<0.001$
  & $(-15.202,\ -13.909)$ \\
\bottomrule
\end{tabular}
\end{table}

\subsubsection{Frailty and non-frailty comparison}

Frailty models are used to account for unobserved heterogeneity across
children~\cite{rabe2008multilevel,yamaguchi1991}. Comparison
of the frailty and non-frailty models shows that the main coefficient
patterns remain stable, but estimated magnitudes increase once latent
heterogeneity is incorporated. 
A no frailty model in Table \ref{tab:nf} is compared to a frailty model in Table \ref{tab:f}. 
The frailty model is preferred. The likelihood-ratio test for latent
heterogeneity is highly significant ($\rho = 0.63$, $p<0.001$),
indicating that unobserved heterogeneity accounts for a substantial
fraction of the variance. The variance estimate $\sigma_u = 2.38$
implies marked between-child heterogeneity in mortality risk beyond
that explained by the observed covariates. Comparing
Tables~\ref{tab:nf} and~\ref{tab:f}, the non-frailty model attenuates
the main coefficients  -  most clearly for maternal death and refugee
status. Once heterogeneity is accounted for, maternal death remains
the dominant mortality correlate, refugee status remains positively
associated with child death, and the early-life concentration of
mortality risk is preserved.

\begin{table}
\centering
\caption{Non-frailty model for child death.}
\label{tab:nf}
\small
\begin{tabular}{llccc}
\toprule
 & Variable & Coefficient & $p$-value & 95\% CI \\
\midrule
\multirow{4}{*}{Analysis age (base: 0)}
  & 1 &  $0.123$ & $0.215$ & $(-0.071,\ 0.318)$ \\
  & 2 & $-0.579$ & $<0.001$ & $(-0.825,\ -0.333)$ \\
  & 3 & $-1.607$ & $<0.001$ & $(-1.936,\ -1.277)$ \\
  & 4 & $-1.963$ & $<0.001$ & $(-2.358,\ -1.569)$ \\
\midrule
\multirow{3}{*}{Mother}
  & Migration  & $-0.043$ & $0.787$ & $(-0.360,\ 0.273)$ \\
  & Status (dead vs alive) & $3.269$ & $<0.001$
    & $(2.433,\ 4.104)$ \\
  & Refugee status & $0.333$ & $<0.001$ & $(0.167,\ 0.499)$ \\
\midrule
\multirow{4}{*}{Mother age (base: 15--24)}
  & 25--29 & $0.290$ & $0.023$ & $(0.051,\ 0.539)$ \\
  & 30--34 & $0.291$ & $0.030$ & $(0.028,\ 0.555)$ \\
  & 35--39 & $0.164$ & $0.276$ & $(-0.131,\ 0.461)$ \\
  & 40--59 & $0.163$ & $0.271$ & $(-0.127,\ 0.455)$ \\
\midrule
& Gender & $-0.021$ & $0.796$ & $(-0.181,\ 0.138)$ \\
\midrule
\multirow{2}{*}{Succeeding sibling}
  & Birth & $0.492$ & $0.056$ & $(-0.013,\ 0.998)$ \\
  & Death & $-13.171$ & $<0.001$
    & $(-22.008,\ -4.335)$ \\
\midrule
Preceding sibling & Death & $-14.199$ & $<0.001$
  & $(-14.912,\ -13.486)$ \\
\bottomrule
\end{tabular}
\end{table}
  \begin{table}
\centering
\caption{Frailty model for child death (preferred specification), with $\rho = 0.63$ and $p < 0.001$, indicating significant unobserved heterogeneity.}
\label{tab:f}
\small
\begin{tabular}{llccc}
\toprule
 & Variable & Coefficient & $p$-value & 95\% CI \\
\midrule
\multirow{4}{*}{Analysis age (base: 0)}
  & 1 &  $0.333$ & $0.012$ & $(-0.072,\ 0.595)$ \\
  & 2 & $-0.353$ & $0.024$ & $(-0.660,\ -0.047)$ \\
  & 3 & $-1.437$ & $<0.001$ & $(-1.826,\ -1.049)$ \\
  & 4 & $-1.886$ & $<0.001$ & $(-2.325,\ -1.446)$ \\
\midrule
\multirow{3}{*}{Mother}
  & Migration  & $-0.089$ & $0.619$ & $(-0.443,\ 0.264)$ \\
  & Status (dead vs alive) & $4.349$ & $<0.001$
    & $(2.989,\ 5.708)$ \\
  & Refugee status & $0.419$ & $<0.001$ & $(0.201,\ 0.637)$ \\
\midrule
\multirow{4}{*}{Mother age (base: 15--24)}
  & 25--29 & $0.367$ & $0.022$ & $(0.053,\ 0.681)$ \\
  & 30--34 & $0.390$ & $0.021$ & $(0.059,\ 0.721)$ \\
  & 35--39 & $0.241$ & $0.189$ & $(-0.118,\ 0.600)$ \\
  & 40--59 & $0.225$ & $0.212$ & $(-0.128,\ 0.580)$ \\
\midrule
& Gender & $-0.006$ & $0.949$ & $(-0.214,\ 0.200)$ \\
\midrule
Succeeding sibling & Birth & $0.533$ & $0.070$
  & $(-0.044,\ 1.112)$ \\
\midrule
\multicolumn{5}{l}{$\rho = 0.63$,\ $p<0.001$;\quad $\sigma_u = 2.38$}\\
\bottomrule
\end{tabular}
\end{table}

\subsubsection{Final EHA model}
\label{sec:finalEHA}

The final EHA interpretation is based on the frailty model, excluding
sibling death variables from substantive conclusions. Maternal death
is the strongest predictor of child death, maternal refugee status
remains positively associated with mortality, and the hazard declines
after the earliest years of life. Succeeding sibling birth shows weak
but plausible positive association, particularly in the trailing
specification, whereas maternal migration and child gender do not
emerge as robust predictors. 
Model adequacy is assessed via a link test (Table~\ref{tab:misspec}).
The significant \texttt{hat} term and insignificant \texttt{hatsq}
term ($p=0.971$) confirm no evidence of misspecification. 
 Together, the EHA results identify a stable core of mortality
correlates: maternal death, maternal refugee status, and the
concentration of mortality risk early in life, with weaker evidence
for succeeding sibling birth and little evidence for child gender or
temporary maternal migration as independent predictors.

\begin{table}
\centering
\caption{Link test for the frailty model.}
\label{tab:misspec}
\begin{tabular}{lccc}
\toprule
Variable & Coefficient & $p$-value & 95\% CI \\
\midrule
\texttt{hat}   & $1.010$   & $0.001$ & $(0.435,\ 1.585)$ \\
\texttt{hatsq} & $0.00115$ & $0.971$ & $(-0.060,\ 0.627)$ \\
\bottomrule
\end{tabular}
\end{table}

 



\subsection{OT Results}
\label{sec:fullorbit}

The EHA results identify a stable subset of variables associated with child death within a well-specified hazard model. In particular, maternal death and early-life age structure emerge as statistically
significant components of the mortality process (Table~\ref{tab:f}).
Maternal migration does not reach significance in any EHA
specification ($p > 0.6$, Tables~\ref{tab:seffect}--\ref{tab:f}).
  However, the EHA formulation operates on an aggregated representation
of the data. The EHA model estimates $\mathbb{P}(CS_t = 0 \mid X_t)$,
projecting the full system onto a scalar hazard. In doing so, the
joint configuration of variables and the transition structure between
states are not explicitly represented. 
 We therefore ask: does a reduced state-space representation exist 
that preserves the significant transition dynamics identified by EHA
while retaining the trajectory structure of the full system?  The orbit
analysis that follows shows that such a representation exists and is
induced by a three-variable subsystem governing the principal
transition mechanisms associated with mortality.
 Rather than modelling the probability of
death, we examine how observed trajectories move through an exact
state space and whether the significant statistical findings identified
by EHA are reflected in the transition structure of the data.
 The orbit analysis reveals that trajectories occupy a highly
constrained subset of the full state space and that transitions into
child death states occur through a limited number of significant state
transitions driven primarily by maternal variables. This structure is read directly from the observed trajectories. 

\begin{table} 
\caption{The 12 OT variables and their abbreviations.} 
\label{tbl:vars}
\centering
\small
\begin{tabular}{lll}
\toprule
Index & Abbreviation & Meaning \\
\midrule
$q_0$ & CS  & Child status (alive/dead) \\
$q_1$ & MS  & Mother status (alive/dead) \\
$q_2$ & MM  & Mother migration (in/out) \\
$q_3$ & RS  & Refugee status \\
$q_4$ & MAG2 & Mother age at birth ($\geq 25$ fit) \\
$q_5$ & Gender & Child gender \\
$q_6$ & SSB & Succeeding sibling birth \\
$q_7$ & SSD & Succeeding sibling death \\
$q_8$ & PSD & Preceding sibling death \\
$q_9$ & TSB & Trailing sibling birth effect \\
$q_{10}$ & TSD & Trailing sibling death effect \\
$q_{11}$ & TPD & Trailing preceding sibling death effect \\
\bottomrule
\end{tabular}
\end{table}

Table \ref{tbl:vars} gives the 12 OT variables,  their abbreviations, and $q$-index assignments. The unfavourable/favourable 
(0/1) coding follows Table~\ref{tab:otvars}. 
 The frequency of change of each variable is computed over the full
dataset. Fundamentally unchanging variables (e.g., child gender) are
ordered first; changing variables are then arranged in ascending order
of their population-wide frequency of change across the 31,081
children. The frequencies of change of the 12 variables are
\begin{center}
    
$q_3 (RS)\!:\!0,\quad q_4 (MAG2)\!:\!0,\quad q_5 (Gender)\!:\!0,\quad q_{11}(TPD)\!:\!19,\quad
q_{10}(TSD)\!:\!22,\quad q_1(MS)\!:\!34,\quad$ \\
$q_7(SSD)\!:\!33,\quad q_8(PSD)\!:\!38,
\quad q_0(CS)\!:\!340,\quad q_9(TSB)\!:\!615,\quad q_6(SSB)\!:\!1246,\quad
q_2(MM)\!:\!4182
$
\end{center}
    The
most stable states (zero frequencies of change) are 
$q_3:$RS, $q_4:$MAG2, and $q_5:$gender. 
Notably, 
 mother
death is rare  while mother migration is  most
frequently changing.   This contrast  -  a
variable that is structurally rare in the population but dominant
in EHA, against a variable that is structurally frequent but
non-significant in EHA  -  motivates the analysis that follows.

\begin{figure}[H]
\centering
\begin{subfigure}[t]{0.51\textwidth}
    \centering
        \includegraphics[height=6cm]{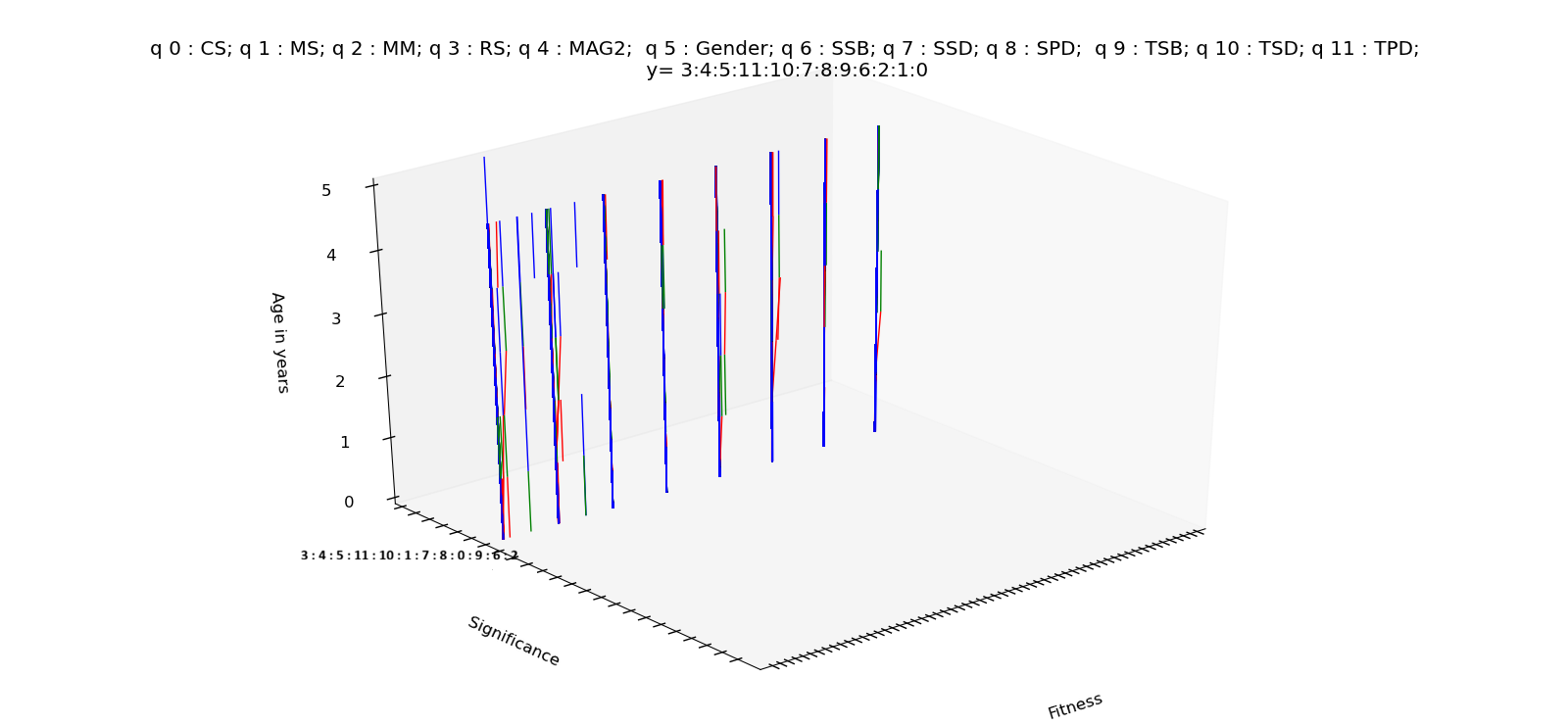}
    \caption{}
    \label{fig:S12_transitions}
\end{subfigure}
\hfill
\begin{subfigure}[t]{0.40\textwidth}
    \centering
     \includegraphics[height=6cm]{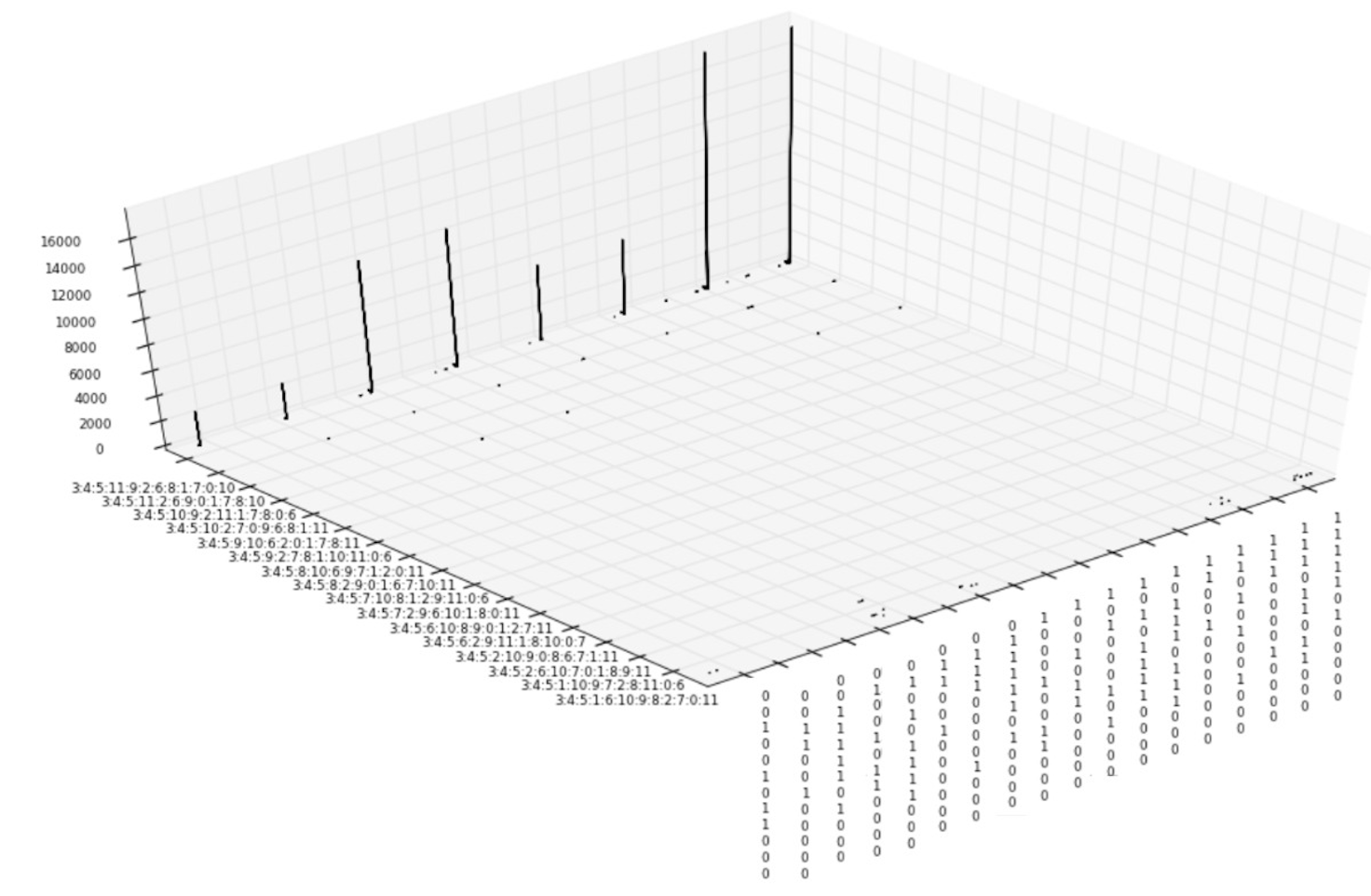}
    \caption{}
       \label{s12hist}
\end{subfigure}
\caption{ (a) Transitions with age over $S_{12}$ for all 31,081 children.   (b) Histogram of state visits in $S_{12}$.}
\label{fig:S12_combined}
\end{figure}

Figure~\ref{fig:S12_combined}(a) shows all observed orbit-state transitions in $S_{12}$   over time (age) -   red lines indicate a move to a less fit state, green to a fitter 
state, and blue for unchanged fitness. 
  These blue state transitions in which two variables change value simultaneously result in unchanged fitness  but  change in significance order.  This is a structural feature that Markov analysis cannot detect.  Markov projects states in vertical edges  onto a single significance level, so these vertical edges are invisible to it. OT captures them exactly. 
 The left half of $S_{12}$ ($q_3=0$) localises Mozambican refugees while   the right half ($q_3=1$) localises South African, with 
child death states identified by $q_0=0$. The histogram Figure~\ref{fig:S12_combined}(b) shows that 
population state-visits concentrate  around significance associated to eight variables $y=3:4:5:11:10:1:7:8:****$, where three are invariant ($q_3$ (refugee status), 
$q_4$ (mother age group), and $q_5$ (gender)).  
The most frequently
occupied state  $$
(1101111111111,\; 3\!:\!4\!:\!5\!:\!11\!:\!10\!:\!1\!:\!7\!:\!8
\!:\!0\!:\!9\!:\!6\!:\!2)
$$
   with all favourable  values for the variables except for Gender (0=Boy)  has 17,805 visits. 
   Refugee status partitions the 
population into Mozambican (left four clusters) and South African 
(right four clusters) subgroups, 
with mother age group and child gender producing 
further subdivisions within each cluster. Table \ref{tab:s12subpops}  gives the description of the  clustering according to the first  three stable variables.  

\begin{table}[H]
\centering
\caption{Clustering according to stable variables in $S_{12}$.
The four SA and four Moz  clusters are defined by the three
stable variables $q_3$ (RS), $q_4$ (MAG2), $q_5$ (Gender).}
\label{tab:s12subpops}
\begin{tabular}{cccc}
\toprule
 $y=3\!:\!4\!:\!5\!:\!*\cdots*$  SA  group   & & \\
\midrule
  MAG $<25$ & MAG $<25$ & MAG $>25$ & MAG $>25$ \\
  Girl & Boy & Girl & Boy \\
\midrule
  $x=101{*\cdots*}$ & $x=100{*\cdots*}$ & $x=111{*\cdots*}$
   & $x=110{*\cdots*}$ \\
\midrule $y=3\!:\!4\!:\!5\!:\!*\cdots*$   Moz   group  & &\\
\midrule
  MAG $<25$ & MAG $<25$ & MAG $>25$ & MAG $>25$ \\
  Girl & Boy & Girl & Boy \\
\midrule
  $x=001{*\cdots*}$ & $x=000{*\cdots*}$ & $x=011{*\cdots*}$
   & $x=010{*\cdots*}$ \\
\bottomrule
\end{tabular}
\end{table}

\begin{figure}
\centering
\includegraphics[width=\textwidth]{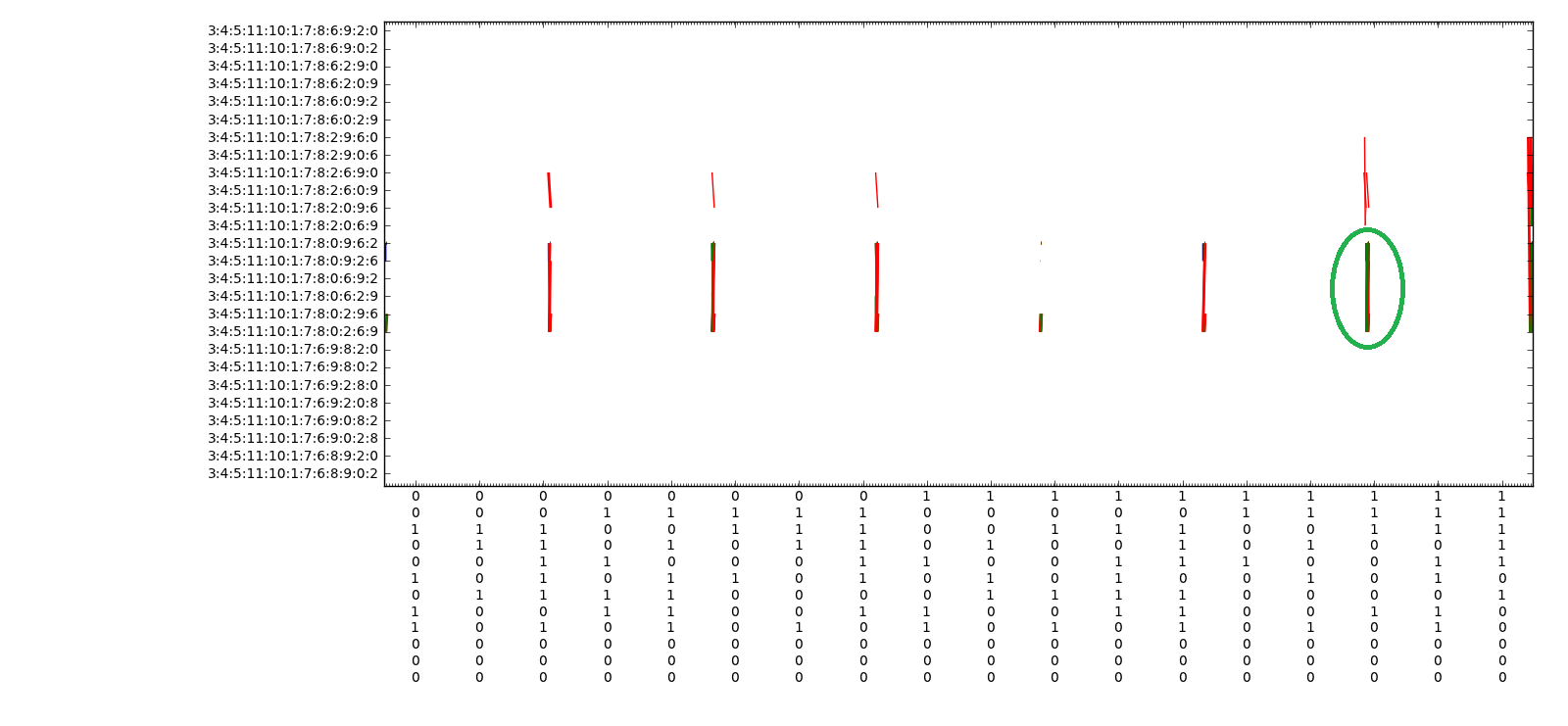}
\caption{Orbits in a subset $R_1\subset S_{12}$ defined by fixing the first seven leading variables
$q_3$(RS), $q_4$(MAG2), $q_5$(Gender), $q_{11}$(TPD),  $q_{10}$(TSD),  $q_1$(MS),  $q_7$(SSD).  At this scale, transitions remain largely overlapping.
The circled region identifies the further subset $R_2\subset
R_1$, magnified in Figure~\ref{fig:S12all2d7a}.} 
\label{fig:S12all2d7b}
\end{figure}

  All transitions are concentrated into a small
subspace for leading variables $q_3, q_4, q_5$.  Transitions hidden in the full data are rich. 
 There is considerable
activity into and from the eight dominant clusters.
 This detail is unresolved at the full scale and
necessitates zooming. 
 At the scale of the full visualisation, most transitions appear as
near-vertical lines  -  indistinguishable in direction and destination (see orbits in subset $R_1$, Figure \ref{fig:S12all2d7b}). 
This is not sparsity in the data but a resolution artefact: the
$y$-axis encodes $12!$ possible variable orderings, so two states
differing only in the least-significant positions of the significance
sequence map to visually adjacent points at full scale. Transition
structure exists but is hidden at this resolution.  To recover it, we restrict attention to the subset
$$
R_1 = \bigl\{(x,y)\in S_{12} : 
y=3:4:5:11:10:1:7:*****\bigr\},
$$
defined by fixing the  first  seven  variables. 
  Fixing these  coordinates
reduces the effective dimension of the fitness axis from $2^{12}$ to
$2^5=32$ fitness configurations, and the significance axis from $12!$ to
$5!$ orderings.
  Of the $5!$ 
   theoretically possible significance orderings within
$R_1$, only 17 are occupied by the observed trajectories, reflecting
the strong persistence of the  fixed variables. Within $R_1$,
transitions between states become geometrically separable: the
$y$-axis now encodes only the orderings of the five remaining
variables, so changes in maternal status ($q_1$) and maternal
migration ($q_2$) produce visually distinguishable displacements.

\begin{figure}
\centering
\includegraphics[width=\textwidth]{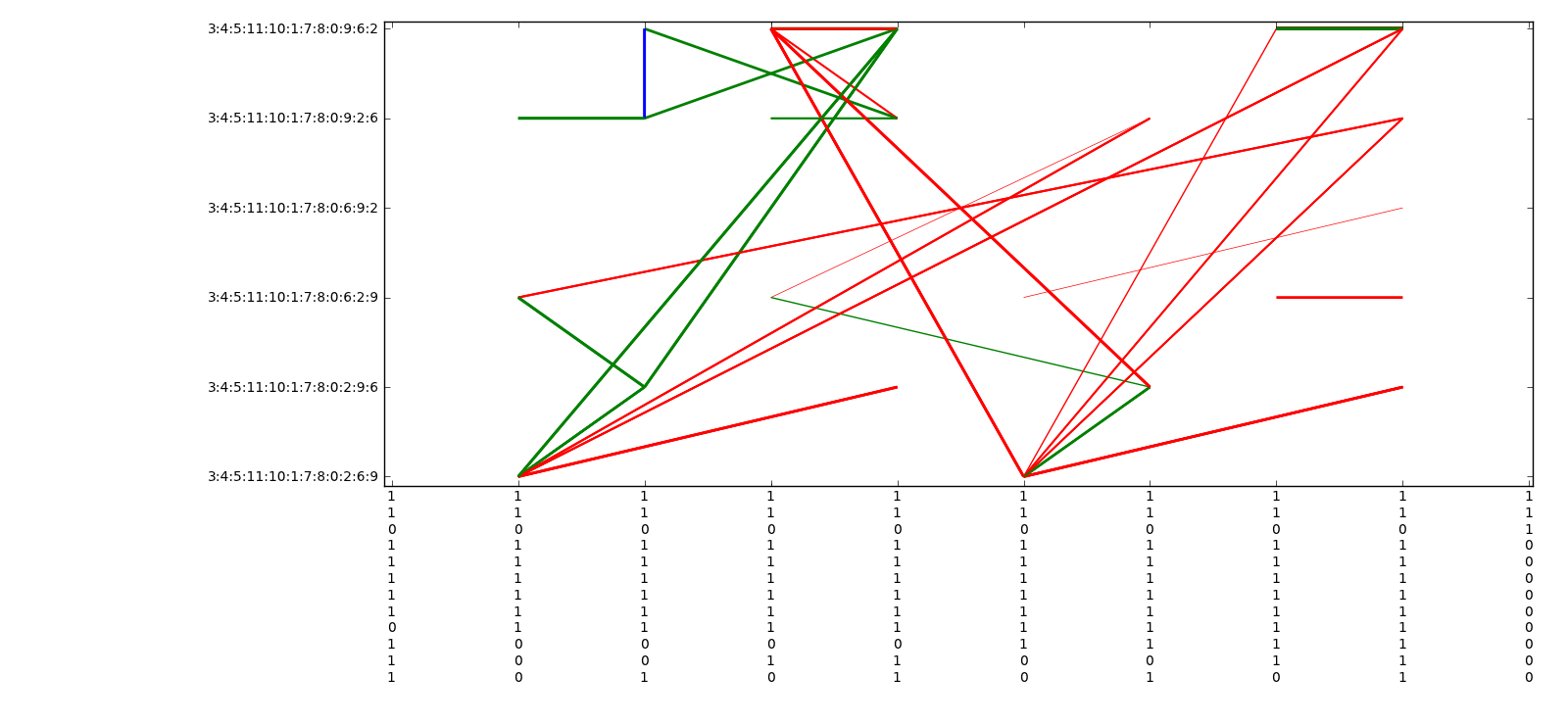}
\caption{The circled region $R_2\subset R_1$
from Figure~\ref{fig:S12all2d7b}, where the first nine variables are fixed,  all with value 1  except for $q_5$(=0 (male)).} 
\label{fig:S12all2d7a}
\end{figure}

Figure \ref{fig:S12all2d7a} shows the zoomed circled region 
from Figure~\ref{fig:S12all2d7b} 
which identifies a
further subset
$$R_2 = \bigl\{(x,y)\in R_1 : x = 110111111a_1a_2a_3,\;
y = 3\!:\!4\!:\!5\!:\!11\!:\!10\!:\!1\!:\!7\!:\!8\!:\!0
\!:\!b_1\!:\!b_2\!:\!b_3 \bigr\},$$ 
where $a_1,a_2,a_3\in\{0,1\}$ and $(b_1,b_2,b_3)$ is a permutation
of $(2,6,9)$, obtained by additionally fixing the first nine variables
and allowing only $q_2$ (MM), $q_6$ (SSB), and $q_9$ (TSB) to vary.
Within $R_2$, individual transition directions are fully resolved: red
lines indicate fitness-decreasing transitions and green lines indicate
fitness-increasing transitions.  
  Since
$q_0=1$ (child alive) and $q_1=1$ (mother alive) are among the
fixed variables, no child or mother death occurs within $R_2$. 
The visible dynamics are entirely those of the three free
variables. 
 
\begin{figure}[H]
\centering
\includegraphics[width=0.8\textwidth]{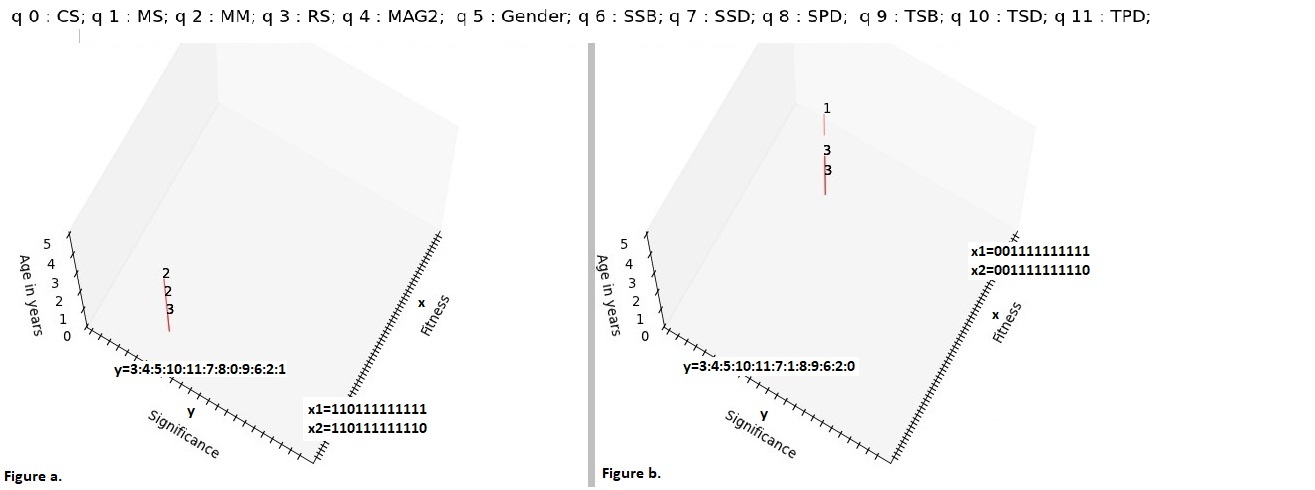}
\caption{Microclusters in $S_{12}\times\text{age}$. Left panel:
Mozambican mother deaths with child alive (rare events). Right panel:
co-occurring mother and child deaths (a small number of cases). }
\label{mcrtrns}
\end{figure}

A microcluster in $S_n$ is a localised set of states 
characterised by low occupancy relative to the dominant 
population clusters and by the presence of at least one 
observed transition into an absorbing state. 
  Figure~\ref{mcrtrns} shows microclusters 
involving mother death and child death in $S_{12}\times$ 
age: the first panel shows Mozambican mother deaths with 
the child alive, and the second shows a small number of 
mother deaths co-occurring with child deaths. The specific 
families involved could in principle be identified and 
referred for individual attention from health professionals. 
The small size of both clusters confirms that simultaneous 
mother and child death is uncommon enough that individual 
rather than aggregate intervention is appropriate.

A special case arises for children observed at only a single age,
termed \textit{point orbits} in OT: these contribute to state-space
occupancy counts but generate no transitions, and correspond
structurally to right-censored observations in EHA.  Point orbits, which carry no transition information, are
treated separately from transition-based analysis. 
   Child-death states ($CS=0$) are absorbing with respect to
child status (once entered, no trajectory can leave via a
change in $CS$) but  not fully absorbing in the Markov
sense if other variables 
 remain free to change after the child's death.
The only fully absorbing states in $S_3$ are those in which
both $CS=0$ and $MS=0$ simultaneously, from which no variable
can transition. Throughout this paper, child-death states are
referred to as \textit{terminal} states  
whether or not other variables
change after the death event. Fully absorbing states,  where both child and mother are dead,  
are labeled \textit{absorbing} in the
strict Markov sense. ~\cite{Norris1997,Kemeny1976}

 Before presenting the reduced orbit system, we note a close 
structural resemblance to both Markov chain 
analysis~\cite{Norris1997,Kemeny1976,SeaWisk23} and multistate 
survival models~\cite{Putter2007}, which similarly represent 
individuals as moving between a set of defined states over time.
The key distinction is that both Markov and multistate frameworks 
fix the state space a priori and estimate transition probabilities 
parametrically, whereas OT constructs the state space directly 
from the data via the significance ordering, without parametric 
assumptions. In classical Markov
analysis, transitions between states for binary data over $n$ questions
are discovered and occupation numbers of states as a fraction of total
occupation numbers define transition probabilities. The key difference
is that Markov analysis fixes the question order and thereby projects
all trajectories onto a single significance level, collapsing $12!-1$
distinct informational dimensions into one. OT retains all of them,
revealing maximum information because it exactly visualises the data
rather than projecting it. 
 Markov analysis applied to the full $S_{12}$ transition data 
would lose all of the significance-axis structure that OT preserves.  It is also the case that it is not then a sharp tool for exposing inhomogeneities in survey data. We add that transition matrices and probabilities do follow in OT. 
They give a powerful computational technique for discovery of dominant 
orbits in the population. But even here the a priori enormous size of the 
state space $S_n$ presents feasibility limits. We have elected to explore 
visual properties in this paper where exact clustering properties and 
dimension reduction give intuitively direct contact with EHA.

\subsubsection{Structural organisation of trajectories}
\label{sec:struct}

Transitions into child death states are not uniformly distributed across the state space. Instead,
they occur along a small number of significant transitions
characterised by changes in maternal variables, particularly  
maternal migration. A further structural feature is the presence of small-capacity states.
High-density states concentrate the majority of transitions and are
most directly relevant for public health intervention. Singleton or
near-singleton child death states reflect rare, individually variable
circumstances (e.g., those associated with sibling death or other
low-prevalence configurations). 
 These correspond to the sparse-exposure configurations that produce unstable coefficients in EHA.  
   Such microclusters are not statistical artefacts but genuine 
features of the data - they represent children whose trajectories  are structurally distinct from the majority and who are invisible 
to aggregate-level analysis. In $S_{12}$, child-death transitions  do not appear among the  most frequent transition types. The 
most common transition -  idling in the fully fit state -  occurs  11,442 times, while child-death transitions individually occur at  most a handful of times each. This confirms that mortality events  are structurally rare relative to the dominant persistence dynamics 
of the system, and that their detection requires a method that  preserves exact trajectory structure rather than aggregating over  states.

A natural question is whether the orbit clusters identified in the reduced system are genuine subgroup structures or artifacts of the deterministic variable-ordering scheme underlying OT. Two features of
the analysis support their substantive interpretation. First, the  numbers of child deaths by age taken over clusters aligns directly with 
the statistically significant effects identified in the EHA model, which
makes no use of the OT encoding. This cross-method convergence is not
guaranteed by the OT construction and provides independent empirical
support that variable re-ordering does not invalidate OT. Second, the 
rare microcluster states correspond precisely to the sparse-exposure configurations (e.g., sibling death combinations) that produce unstable 
or implausible coefficients in EHA. The two methods therefore identify
the same structurally anomalous cases through entirely different
mechanisms. Neither feature would be expected if the clusters were
purely encoding artefacts.

\subsubsection{Reduced orbit system $S_3$}
\label{sec:reduced}

The concentration of trajectories and the dominance of maternal
variables motivate the construction of a reduced system. Define the
mapping
$$
\Phi : \{0,1\}^n \rightarrow \{0,1\}^3, \qquad
\Phi(x_t) = (CS_t,\, MS_t,\, MM_t),
$$
where $CS_t$ denotes child status, $MS_t$ maternal status, and $MM_t$
maternal migration at child age $t$.

\begin{figure} 
\centering
\includegraphics[width=0.7\textwidth]{%
 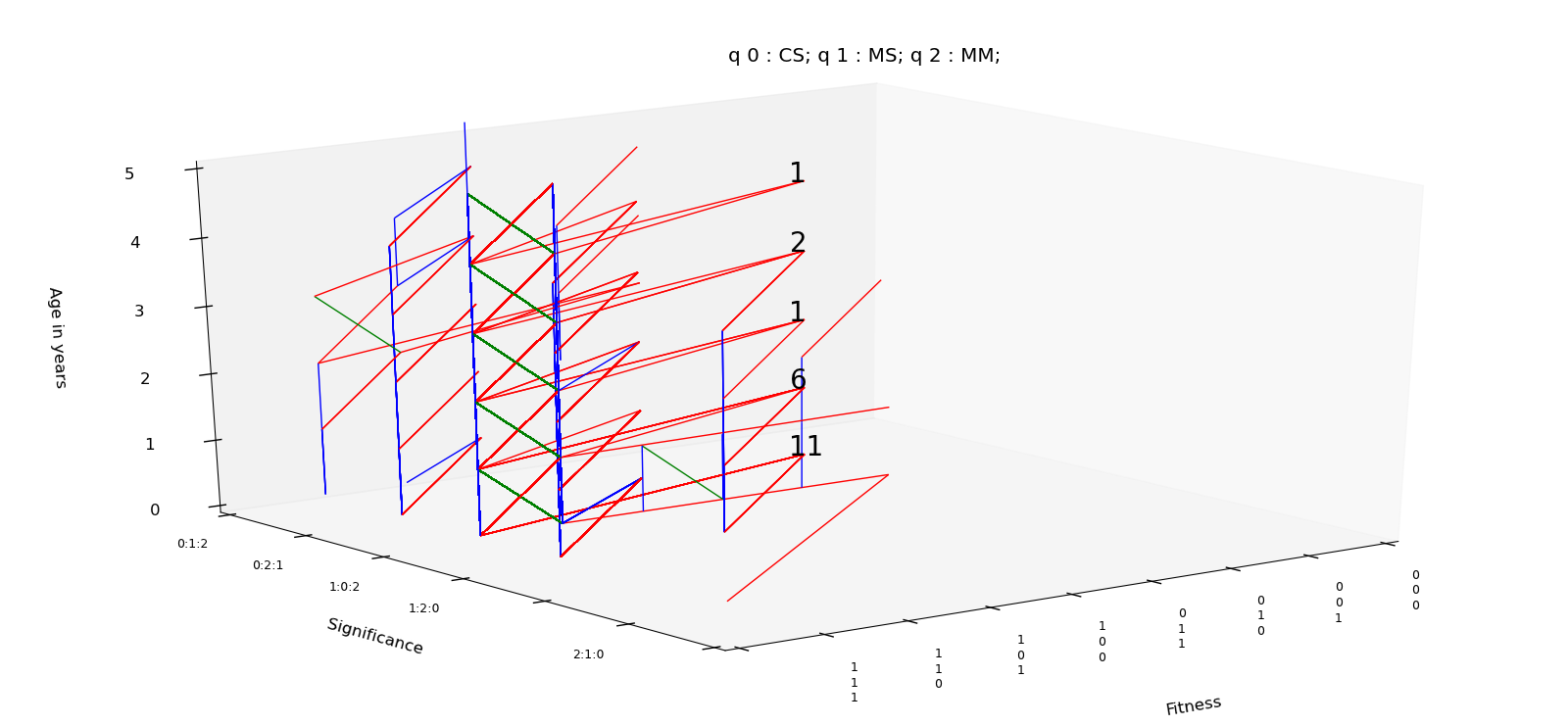}
\caption{Full $S_3$ observed states plotted over  child age in years.}
\label{s3full3d}
\end{figure}

\begin{figure}
\centering
\includegraphics[width=0.7\textwidth]{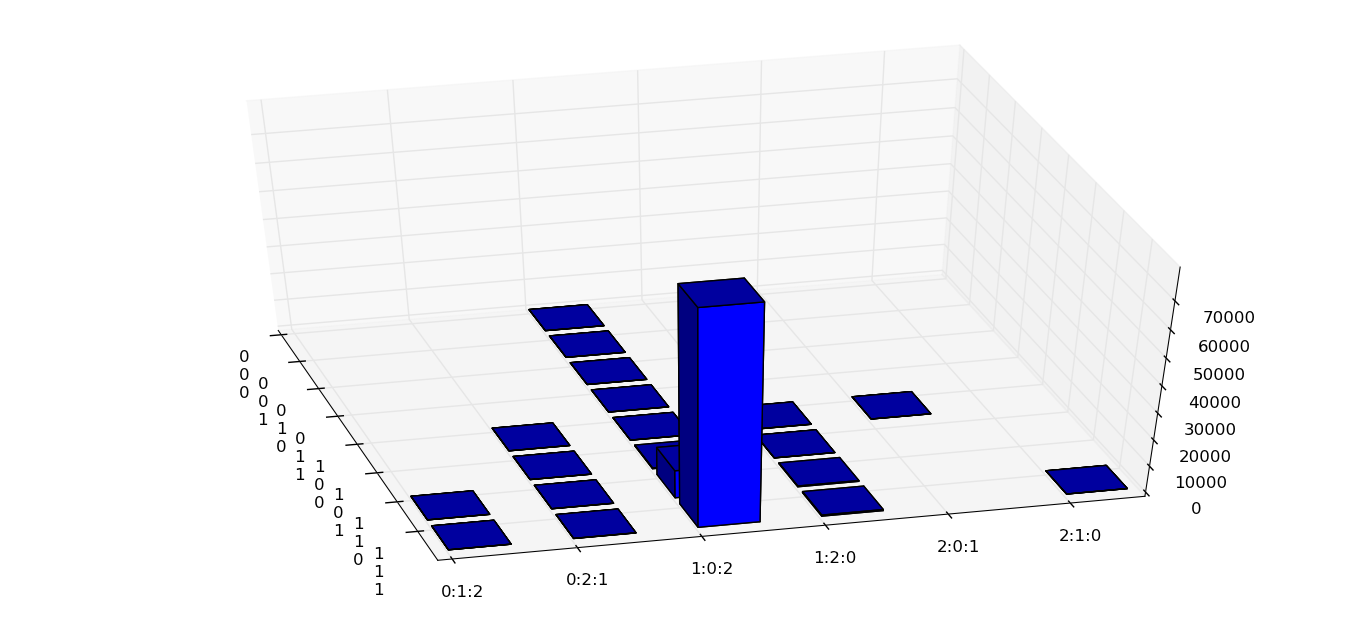}
\caption{$S_3$ state-visit histogram. The most frequently visited
state is $(111,1\!:\!0\!:\!2)$ (child alive, mother
alive, mother present). } 
\label{s3hist}
\end{figure}

\begin{figure}
\centering
\includegraphics[width=0.7\textwidth]{%
 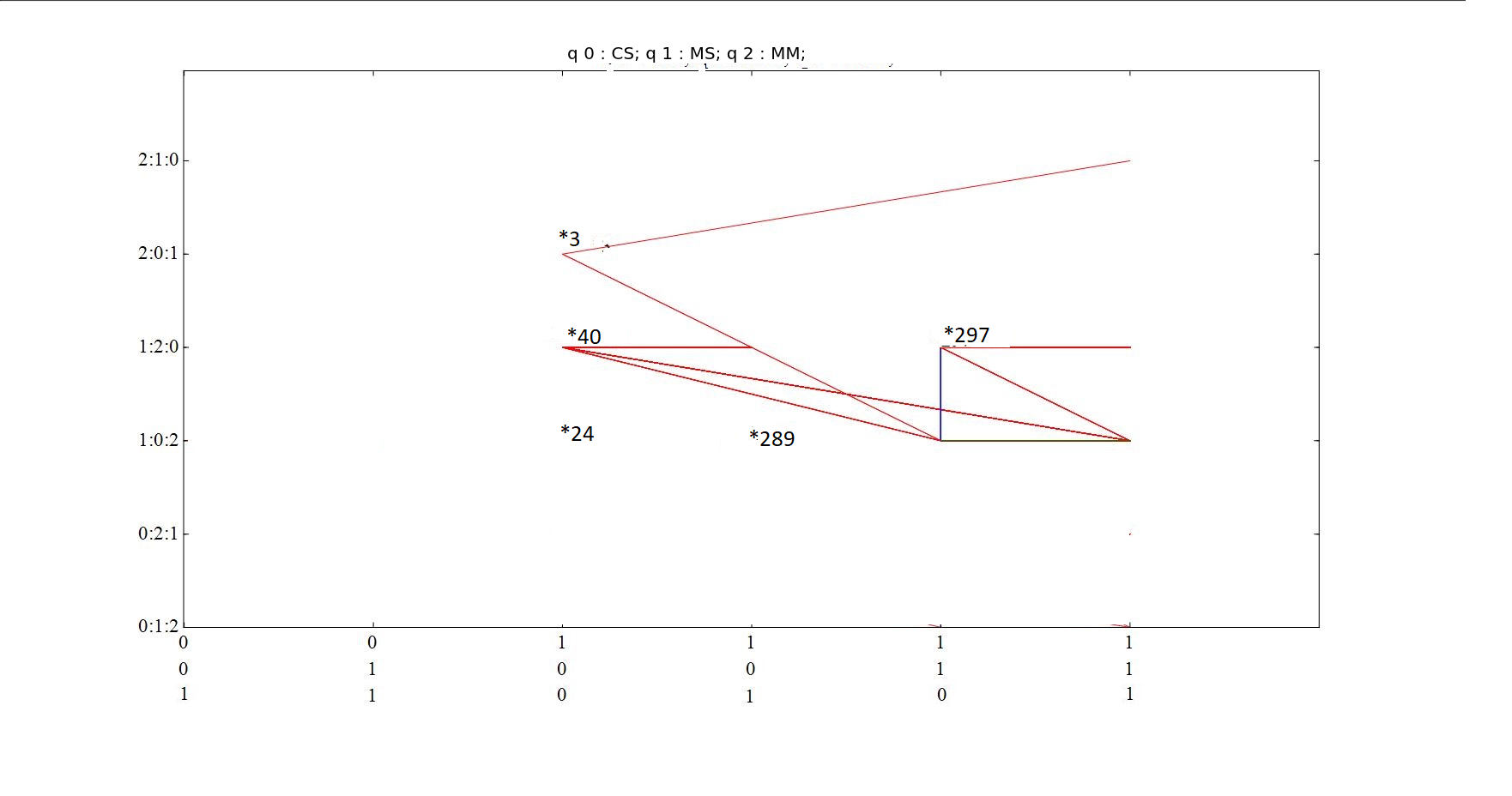}
\caption{Absorbing death states
and maternal-triggered transitions in $S_3$.}
\label{3varsstatesp}
\end{figure}

The choice of variables in $\Phi$ is guided by two criteria: (i)
statistical dominance in the EHA model, and (ii) structural
participation in the orbit representation. 
Recall the frequencies of change of the three variables 
$$
freq(q_1)\!:\!34,\quad freq(q_0)\!:\!340,\quad freq(q_2)\!:\! 4,182.
$$ 
Mother migration ($q_2$) contributes approximately 
 4,182 of the  
  total change-events, or
approximately 92\% of the change-events in the reduced system. 
  This frequency of structural
participation  -  not mere frequency of change in the raw data  - 
provides quantitative support for retaining MM in $\Phi$, independently
of its EHA non-significance.    Figure \ref{s3full3d} shows the full $S_3$ observed states plotted over age in years.   Absorbing death states and maternal-triggered transitions are
visible. Death frequency decreases with increasing age, consistent
with the EHA age-hazard profile.
 Figure \ref{s3hist}  shows that the most frequently visited state $(111,1:0:2)$  is where both child and mother are alive,  and where mother is present. The dominant population  resides in fit, non-death states.
  Figure \ref{3varsstatesp} shows the  absorbing death states and maternal-triggered transitions. At  the
point-state (x = 101, y = 1 : 0 : 2), there are  289 point orbits, i.e. child deaths in the year of first observation, accumulated over the analysis period.
There are 297 transitions  terminating in (MA, MI,CD)  accumulated over the observation years.
  


In particular, the question is which state containing a living 
child most often precedes child death.
 Maternal migration (MM) is the clearest illustration of this principle.
In EHA, MM is statistically non-significant in every specification
($p>0.6$, Tables~\ref{tab:seffect}--\ref{tab:f}).
  In the orbit
representation, the picture is strikingly different.

Of the 340 observed sequential transitions into child-death 
states (Blocks~I--III, Table~\ref{s3tab:trans}), 297 (87\%) 
terminate in the mother-present child-death configuration 
$(MA, MI, CD)$ -  a structural pattern organised by 
maternal migration status that has no analogue in the EHA  output.   Maternal death appears
as a changing variable in only 2 transitions  -  those in
Block~III where mother and child death occur simultaneously. The
remaining 41 transitions (Block~II) involve child death with the mother
alive and no  change in either MM or MS. A further 313 children
enter child-death states as point orbits with the mother alive
(Block~IV), contributing no observed sequential transition.  Table~\ref{s3tab:trans} shows that child death occurs in both
MM states: 290 transitions from (MA,MI,CA) to (MA,MI,CD) and
37 from (MA,MO,CA) to (MA,MO,CD). Since child death occurs in both mother migration status  (in or out), the conditional probability difference between
them is small, and MM does not reach significance in EHA
($p>0.6$). MM is therefore retained in $\Phi$  because it is structurally pivotal:
it organises 87\% of the observable sequential transitions into
child-death states, a distinction entirely invisible to EHA's marginal
hazard estimate. 
The denominator here is the 340 observed two-state transitions 
(Blocks~I--III of Table~\ref{s3tab:trans}), which excludes the 313 point orbits 
recorded in Block~IV. Point orbits are single-observation trajectories that carry 
no transition information by construction - they record a child's state at one 
age only, with no preceding or succeeding observation from which a sequential 
transition can be identified. Including them in the denominator would conflate 
the absence of observable sequence with evidence of a specific transition pathway, 
which is not meaningful. The 87\% figure therefore describes the proportion of 
children whose path to death was observationally sequential and for whom MM 
organises that sequence. This is a structural-participation argument grounded
in  
 the  transition counts, not a
frequency-of-change argument.
 This contrast is the central structural asymmetry the reduced orbit
system exposes. 
EHA identifies maternal death as the dominant
statistical predictor (coefficient $4.349$, $p<0.001$,
Table~\ref{tab:f}), yet maternal death appears as a changing
variable in only 2 of the 340 observed sequential transitions in $S_3$.
MM, statistically null in EHA, organises 297 of those same 340
transitions. The two findings are not contradictory: maternal death is
so strongly absorbing and so rare (34 population-wide change-events,
Section~\ref{sec:fullorbit}) that it rarely appears as a preceding
state in a year-to-year transition  -  its role is terminal rather than
sequential. OT detects a structurally changing variable that EHA does not. EHA detects a statistically dominant variable 
whose structural role is expressed through the absorbing 
geometry of the state space rather than through observed 
two-state transitions -  a consequence of maternal death's 
terminal rather than sequential role, and of the short 
time series available per child.  
\begin{table}
\centering
\caption{Significant transitions in $S_3$ terminating in child-death
states. States are decoded using the $y$-ordering: for
$y=y_0\!:\!y_1\!:\!y_2$, the bits of $x$ are read left to right as
variables $y_0, y_1, y_2$ respectively, where $0=$CS, $1=$MS,
$2=$MM. Abbreviations: CA = child alive, CD = child dead,
MA = mother alive, MD = mother dead, MI = mother in, MO = mother out.}
\label{s3tab:trans}
\small
\begin{tabular}{llcc}
\toprule
State 1 & State 2 & Meaning & Count \\
\midrule
\multicolumn{4}{l}{\textit{Block I: transitions terminating in
$(MA,MI,CD)$}} \\
$(x=111,y=1\!:\!2\!:\!0)$ & $(x=110,y=1\!:\!2\!:\!0)$
  & (MA,MI,CA)$\to$(MA,MI,CD) & 290 \\
$(x=111,y=1\!:\!0\!:\!2)$ & $(x=110,y=1\!:\!2\!:\!0)$
  & (MA,CA,MI)$\to$(MA,MI,CD) & 4 \\
$(x=110,y=1\!:\!0\!:\!2)$ & $(x=110,y=1\!:\!2\!:\!0)$
  & (MA,CA,MO)$\to$(MA,MI,CD) & 3 \\
\multicolumn{3}{r}{\textit{Subtotal}} & \textbf{297} \\
\midrule
\multicolumn{4}{l}{\textit{Block II: transitions terminating in
$(MA,MO,CD)$}} \\
$(x=101,y=1\!:\!2\!:\!0)$ & $(x=100,y=1\!:\!2\!:\!0)$
  & (MA,MO,CA)$\to$(MA,MO,CD) & 37 \\
$(x=111,y=1\!:\!0\!:\!2)$ & $(x=100,y=1\!:\!2\!:\!0)$
  & (MA,CA,MI)$\to$(MA,MO,CD) & 2 \\
$(x=110,y=1\!:\!0\!:\!2)$ & $(x=100,y=1\!:\!2\!:\!0)$
  & (MA,CA,MO)$\to$(MA,MO,CD) & 1 \\
  $(x=111,y=2\!:\!1\!:\!0)$ & $(x=100,y=1\!:\!2\!:\!0)$
  & (MI,MA,CA)$\to$(MA,MO,CD) & 1 \\
\multicolumn{3}{r}{\textit{Subtotal}} & \textbf{41} \\
\midrule
\multicolumn{4}{l}{\textit{Block III: simultaneous mother and child
death}} \\
$(x=110,y=1\!:\!0\!:\!2)$ & $(x=100,y=2\!:\!0\!:\!1)$
  & (MA,CA,MO)$\to$(MI,CD,MD) & 2 \\
\midrule
\multicolumn{4}{l}{\textit{Block IV: absorbing single-state death
occupations (point orbits)}} \\
$(x=100,y=1\!:\!0\!:\!2)$ &  -  & (MA,CD,MO) & 24 \\
$(x=101,y=1\!:\!0\!:\!2)$ &  -  & (MA,CD,MI) & 289 \\
\multicolumn{3}{r}{\textit{Subtotal}} & \textbf{313} \\
\bottomrule
\end{tabular}
\end{table} 

\begin{figure}
\centering
\begin{subfigure}[t]{\textwidth}
    \centering
    \includegraphics[height=5cm]{%
       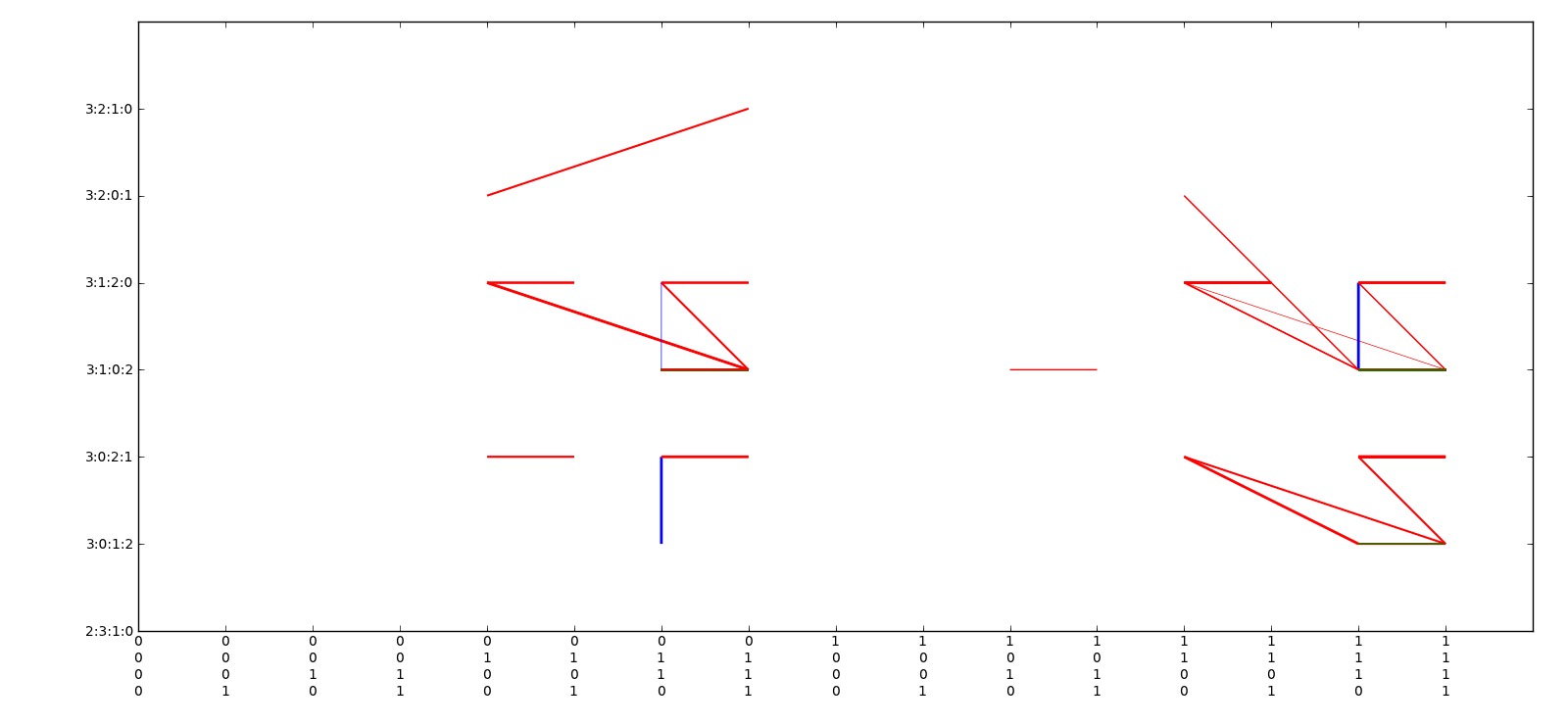}
    \caption{}
    \label{fig:s4trans2d}
\end{subfigure}

\vspace{0.5em}

\begin{subfigure}[t]{0.49\textwidth}
    \centering
    \includegraphics[height=6cm]{%
       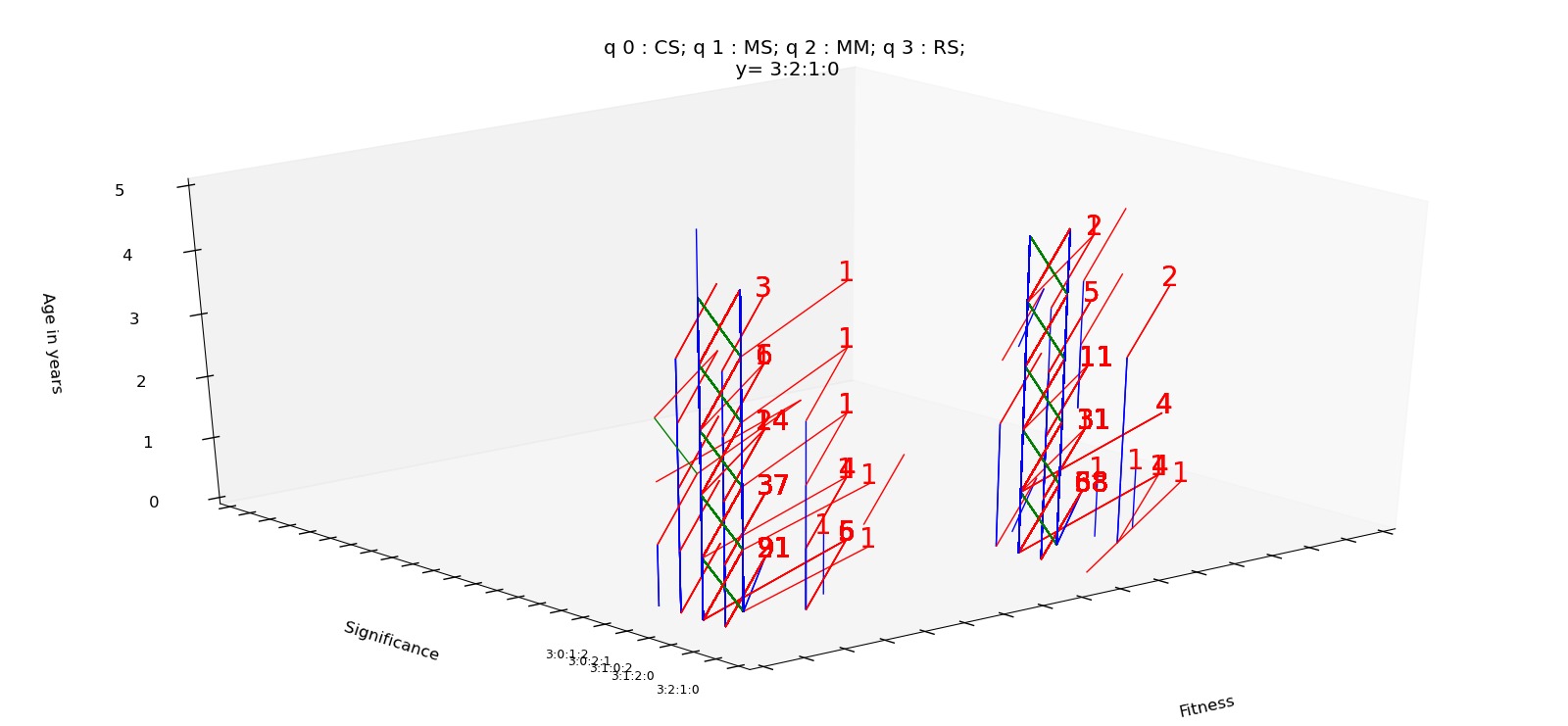}
   \caption{}
    \label{fig:s4trans3d}
\end{subfigure}
\hfill
\begin{subfigure}[t]{0.49\textwidth}
    \centering
    \includegraphics[height=6cm]{%
       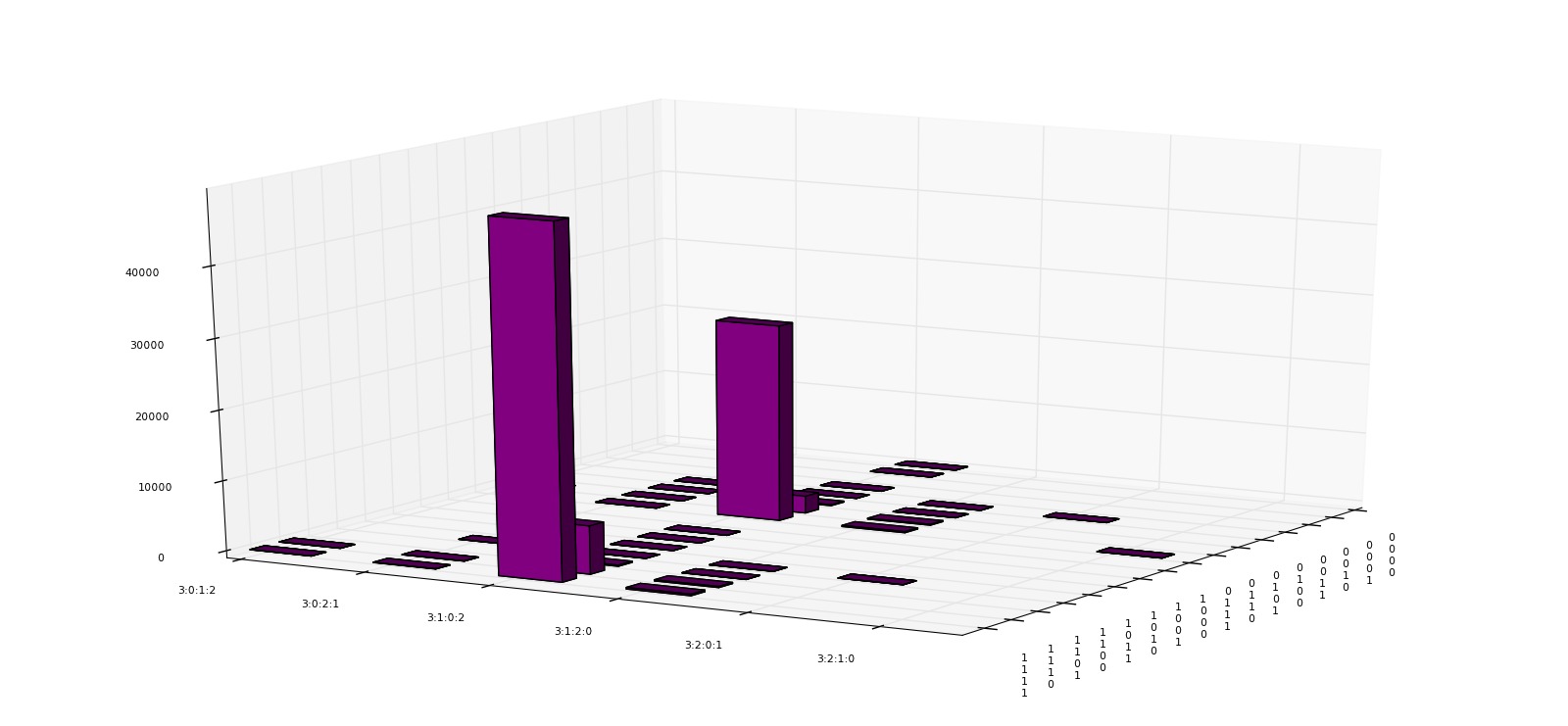}
    \caption{}
    \label{fig:s4histo}
\end{subfigure}

\caption{Four-variable orbit system $S_4$ under
$\Phi_4=(CS,MS,MM,RS)$.
\textbf{(a)}~Two-dimensional transitions: Mozambican ($RS=0$,
left) and South African ($RS=1$, right) subpopulations share
identical $y$-orderings and transition patterns.
\textbf{(b)}~Three-dimensional transitions over child age: both
clusters show the same structure, with death-entry transitions
concentrated at ages 0--1 and declining through ages 2--4.
\textbf{(c)}~State-visit histogram: two dominant states
($\approx$45,000 and $\approx$33,000 visits) correspond to
the fully fit configurations of the Mozambican and South
African subpopulations respectively.}
\label{fig:s4combined}
\end{figure}

Maternal refugee status (RS) illustrates the converse case. 
RS is statistically significant in EHA (coefficient $0.419$, 
$p<0.001$, Table~\ref{tab:f}), but it is constant for each 
child and therefore generates no transitions. A constant 
coordinate can only stratify the population into fixed 
subgroups (Table~\ref{tab:s12subpops}: Mozambican 
vs.\ South African), not organise trajectory dynamics. Figures~\ref{fig:s4trans2d} and~\ref{fig:s4trans3d} show 
the transition structure under a four-variable reduction 
$\Phi_4=(CS,MS,MM,RS)$, extending $\Phi$ by adding refugee 
status. Two clusters emerge - Mozambican ($RS=0$) and 
South African ($RS=1$). The structural organisation within 
each cluster is identical: the same $y$-orderings, the same 
dominant death-entry pathways, and the same age-concentration 
of transitions appear in both subpopulations. RS therefore 
acts as a fixed partitioning variable -  it separates the 
population into two subgroups of different size but does not 
alter the dynamics within either.
 This confirms the exclusion criterion: RS is 
statistically significant in EHA but structurally inert in 
OT, generating no transitions of its own and contributing 
no additional organisational structure beyond what 
$\Phi=(CS,MS,MM)$ already captures. The reduced system is 
therefore not selected by statistical significance alone, 
but by the ability of variables to generate and organise 
transition structure. Adding RS produces parallel copies 
of the same dynamics rather than new dynamics. 
 The reduction $\Phi$ is not a zoom into $S_{12}$. Zooming into a
subset $R \subset S_{12}$ by fixing $k$ variables restricts
attention to children for whom those $k$ variables take specific
constant values, leaving the remaining $n-k$ variables free. The
resulting subspace has at most $2^{n-k}\cdot(n-k)!$ states, which
for $k=9$ gives the same $2^3\cdot3!=48$ states as $S_3$. But the
two representations are fundamentally different. In $R$, the fixed
variables are constants by restriction -  they carry no transition
information within the subspace. In $S_3$ under $\Phi=(CS,MS,MM)$,
all three variables are free to change; $CS$ and $MS$ are changing
variables whose transitions define the mortality dynamics of
interest. A zoom that fixes $CS$ and $MS$ produces a 48-state space
in which child death and mother death are invisible by construction.
The map $\Phi$ is therefore not a zoom: it is a projection that
retains the mortality-relevant variables as dynamic.

The projection $\Phi$
aggregates across the full population: every child contributes to
$S_3$ regardless of their values on the nine variables outside
$\Phi$. Two children with entirely different refugee status, sibling
configurations, and maternal age occupy the same $S_3$ state if
their $(CS, MS, MM)$ values match. This aggregation is what makes
$S_3$ a genuine reduction rather than a subpopulation analysis, and
what allows the transition counts in Table~\ref{s3tab:trans} to
reflect the full 31,081 children.
The map $\Phi$ is therefore a search for homogeneous dominant associations
in the exact data. EHA identifies maternal death as a hazard in a
heterogeneous dataset; OT identifies maternal migration as the
variable most frequently associated with child death transitions.
We ask whether $(CS, MS, MM)$ together support homogeneity in the
exact data - whether data can be aggregated on common patterns of
transitions,  and whether this yields a structurally coherent
account of the associations among these three variables.

The reduced system is minimal in the following sense. For any
changing variable, 
  removal from $\Phi$ merges states
with genuinely distinct transition histories, destroying structural
information. For any variable whose value is fixed
for every individual, removal is lossless: since no individual
ever transitions on that variable, dropping it collapses the
population into fewer subgroups but generates no false transitions
and erases no transition dynamics.
 Removing $MS$ from $\Phi$ would merge $x=110$ (child alive, mother
alive, mother out) and $x=100$ (child alive, mother dead, mother
out), conflating the 297 Block~I transitions with the 41 Block~II
transitions in Table~\ref{s3tab:trans} -  a distinction the data
clearly support as structurally separate. Removing $MM$ would merge
$x=111$ and $x=101$, eliminating the ability to distinguish
trajectories that reorganise (mother migrates) before death from
those that do not. Both are changing: $MS$ changes for 34 individuals
and $MM$ for 4{,}182 across the observation window
(Section~\ref{sec:fullorbit}). By contrast, $RS$ is invariant - 
no child changes refugee status over the observation period, giving
$\text{freq}(q_3)=0$. Dropping $RS$ from $\Phi$ generates no false
transitions and erases no transition dynamics: any trajectory in
$S_4$ under $\Phi_4=(CS,MS,MM,RS)$ maps to an identical trajectory
in $S_3$ under $\Phi$, with the same transition counts and absorbing
structure. What is lost is only the subpopulation label -  the
ability to distinguish Mozambican from South African children within
any given $(CS,MS,MM)$ state. This is confirmed empirically in
Figure~\ref{fig:s4combined}: the two $S_4$ clusters are congruent
copies of the same transition structure, one per $RS$ value. Adding
$RS$ to $\Phi$ produces two parallel copies of $S_3$ rather than
new dynamics. The three-variable reduction $\Phi=(CS,MS,MM)$ is
therefore minimal: it retains all  transition-generating
variables and excludes all constant ones.

The reduction from $S_{12}$ to $S_3$ is not a random compression.
The observed orbits concentrate along a small number of $y$-levels
determined by the most stable variables. The reduced system
represents the mortality-relevant support of the data, collapsing
configurations that do not change the observed transition structure.
$S_3$ has $|X_3|=2^3=8$ possible binary configurations and
$|Y_3|=3!=6$ possible variable orderings, giving at most 48 states
-  entirely tractable for direct enumeration and transition
analysis without parametric assumptions.
Table~\ref{s3tab:trans} lists the significant transitions in
Figure~\ref{3varsstatesp}, organised into four blocks: (I)
transitions terminating in $(CD,MA,MI)$; (II) transitions
terminating in $(CD,MA,MO)$; (III) transitions terminating in
$(CD,MD,MI)$; and (IV) absorbing single-state death occupations
(point orbits).

\begin{table}
\centering
\caption{Correspondence between the EHA age-hazard profile
(frailty model, Table~\ref{tab:f}) and the age distribution of
significant $S_3$ transitions into child-death states
(Table~\ref{s3tab:trans}). Age 0 is the reference category in EHA
(coefficient $=0$).  OT counts are non-trivial transitions
only (point orbits excluded).} 
\label{tab:age_correspondence}
\begin{tabular}{lccccc}
\toprule
Analysis age & 0 & 1 & 2 & 3 & 4 \\
\midrule
EHA frailty-model age coefficient
  & $0$ (ref.) & $0.333$ & $-0.353$ & $-1.437$ & $-1.886$ \\
$S_3$ transitions into child-death states
  & 194 & 85 & 38 & 22 & 13 \\
\bottomrule
\end{tabular}
\end{table}


    
The transitions in the three dominant variables are extracted and
visualised in Figure~\ref{3varsstatesp} and Table~\ref{s3tab:trans}. Maternal variables dominate the transition structure leading to child death.  
 Both EHA and OT analyses 
concentrate child 
death in early childhood, reflecting the age-dependent hazard
(Table~\ref{tab:age_correspondence}).  Maternal migration, despite
being statistically null in EHA, organises 87\% of the observable
sequential transitions into child-death states
(Table~\ref{s3tab:trans})  -  a finding with no analogue in the EHA
output.  The observed trajectories occupy only a small structured
region of $S_{12}$, determined primarily by maternal migration and
child death transitions, with maternal death present as an absorbing terminal condition rather than a sequential transition driver. The reduced orbit system therefore emerges
as a structural concentration phenomenon rather than a visual
simplification by zooming, with the dominant trajectories of the full system
preserved under $\Phi: S_{12} \to S_3$.  The orbit representation makes explicit the trajectories through
which transitions occur and identifies small clusters associated
with elevated risk. These structural features are not directly
observable in EHA, where inference is based on aggregated effects.
Taken together with the 313 point orbits, the clustering of
trajectories reveals inhomogeneity in the full data  -  a risk for
statistical analysis, but also an aid: matching patterns of
trajectories give permission to aggregate. The reduced orbit system
thereby links statistical inference and state-space representation,
providing complementary statistical and structural views of
early-life mortality.

Figure~\ref{3varsstatesp} reveals three paths to the dominant 
child-death cluster of 297 transitions, each originating from 
a fully fit household state. Within the reduced system, the dominant preceding state to the  297-transition child-death cluster is the fully fit household  configuration $(CS=1, MS=1, MM=1)$, with the blue transition  indicating a maternal return from out-migration immediately 
preceding child death. The dominant path involves no 
change in maternal variables: the child dies while the mother 
remains present and alive. A second path begins with a maternal 
return (mother out to mother in) immediately preceding child 
death. Both paths are consistent with the observation that 
mothers return home when their child is close to 
death~\cite{Clark2007}. Causal interpretation requires caution: 
mother presence may reflect a response to impending child death 
rather than a risk factor. The dominant point orbit (289 cases) 
similarly records child death with mother present from the first 
observed age, providing no sequential information about 
preceding events.
   Child death is most closely associated with mothers who are at home  for the majority of the year.  
 The inductive process in OT is natural: the 
presence of two fully fit states that transition to child 
death demands further data. The natural extension is to 
add survey data or new questions and work backwards in 
time, identifying dominant preceding associations at each 
level. In this way $\Phi$ becomes the foundation for a 
growing state space. 

\section{Comparative Analysis of the Methods}
\label{sec:comparative} 

EHA and OT describe the same mortality process through different but 
compatible lenses. EHA estimates the hazard of child death
as a function of covariates, yielding interpretable coefficients,
significance tests, and model diagnostics. OT visualises trajectories
through the first five years of life, preserving the full
combinatorial structure of binary states and exposing transition
patterns, cluster densities, and rare states that aggregate hazard
models cannot recover. Both methods operate on the same underlying
variables and use the same temporal index of child age  -  in EHA as a
covariate affecting the hazard, 
in OT as the axis along which orbits evolve, with 
death-preceding states marking the locations of elevated hazard.  

The  differences are structural. EHA compresses the data into a
parametric form: the discrete-time logistic model estimates the hazard
as a function of selected covariates, controls for unobserved
heterogeneity through frailty, and evaluates model adequacy through
diagnostic tests such as the link test. OT makes no parametric
assumptions and reveals inhomogeneity, at first glance by clustering. 
 EHA aggregates individuals through parametric modelling while  OT clusters them naturally through the literal geometry of shared state-space properties, without imposing a 
statistical criterion.  
  Two-dimensional capacity diagrams and three-dimensional orbit diagrams
reveal clustering, transitions, and the densities of high-risk states 
directly from the data. Aggregation might 
usefully follow if disjoint clusters associate the same variables. 
 EHA reports variable-level log-odds ratios; OT reports 
labelled transition patterns. Both preserve the full 
variable list. Regarding point orbits,  the presence of a large population on short orbits has a
deeper consequence for EHA. Point orbits cannot signal which variable
is a risk to child survival  -  they carry no transition information
 -  yet they contribute to the denominator of the hazard estimate.
If such orbits are numerous and non-randomly distributed across
covariate configurations, the hazard associated with the remaining
variables may be over-estimated for some configurations and
under-estimated for others. This is a fundamental heterogeneity
constraint on EHA that OT makes visible: point orbits are
structurally distinct from non-trivial orbits and are excluded from
the transition-based analysis in $S_3$. More broadly, OT reveals
that the full dataset is inhomogeneous  -  it contains subpopulations
with distinct trajectory structures (point orbits, short orbits, and
full-length orbits)  -  and that matching on common transition
patterns is a prerequisite for aggregation. The map $\Phi$ is
therefore not merely a projection but a search for homogeneous
dominant associations in the exact data: it asks whether the three
variables $(CS, MS, MM)$ by themselves support homogeneity, whether
data can be aggregated on common transition patterns, and whether
this yields a structurally coherent account of the associations among
maternal death, maternal migration, and child death.


The two methods agree on the importance of maternal variables
while attributing dominance to different mechanisms.  What OT adds 
is structural, including also any \emph{objective} (given 
the data, before statistics is considered) heterogeneity. OT shows
not only that these variables are associated with child death, but
also the sequence-dependent trajectories through which high-risk
states arise, and the existence of characteristically distinct
subgroups whose intervention needs differ. 

Each method has limitations. The number of child deaths is large enough to satisfy the needs of both methods. It is well known that EHA is best used for long time series and that data should theoretically be homogeneous~\cite{Singer2003,BoxSteffensmeier2004,Allison1984}.  Standard guidelines require $p \ll 1$, confidence intervals 
excluding the origin, and frailty checks  (Tables~\ref{tab:f} and~\ref{tab:misspec}). In the case of OT, these concerns of EHA do not apply at all.  OT is not sensitive to survey period variability -  a known 
source of unobserved heterogeneity in 
EHA~\cite{Singer2003,BoxSteffensmeier2004}. Where EHA 
compensates through frailty random effects, OT addresses 
heterogeneity structurally, through the congruence structure 
of observed transitions. 

Consider the uniqueness of OT. First, OT rests on exactly two assumptions: 
the a priori $0/1$
(unfit/fit) coding assignment for each variable, and the swap rule
by which a changing variable is moved to the rightmost (least numerically 
significant) position in the individual's fitness with new  significance ordering. Both are reversibly decoded back to original data and the latter is not corrupted in that sense. Regarding the coding: Table~\ref{s3tab:trans} provides an example. In Block~I, mother alive ($MS=1$) and mother in ($MM=1$)
precedes 290 child deaths; in Block~II, mother alive ($MS=1$) and
mother out ($MM=0$) precedes 37 child deaths. The coding $MM=0$ for mother out (unfit) therefore appears justified by the data, \emph{if} mother at home is interpreted as good for the dying child. 
   In OT for given data the coding merely discovers associations, strictly before any attempt to explain the 
results by supplementing the survey with additional information. Note that 
if indeed $MM$ proves to have been coded ``incorrectly'' (i.e. $MM=1$ assigned to mother in), OT does not fail: the correct interpretation can be understood and carried through on the ``incorrect'' visualisations or, since OT executes quickly, the coding can be corrected and visualisations rerun. The associations are unchanged but now have the correct interpretation. 
  This illustrates a fundamental 
difference from Markov analysis: OT admits relative fitness which facilitates 
unique interpretation, Markov analysis does not because it collapses any $S_n$ 
into a single significance (question order). Consider the swap rule. Placing changing variables at the least significant digits of fitness as a number is equivalent, on the number line, to isolating a
portion of state space with invariant fitness and/or slowly changing values 
at the most significant digits.   
 The reduction from $S_{12}$ to $S_3$ is equivalent to shifting changeable values 
to the numerically significant digits on the left and deleting the invariant insignificant digits. The present construction has the practical advantage 
of making demographically important subpopulations (eg MOZ and SA) directly visible and comparable. Where two variables change
simultaneously on a single transition, their relative significance
positions are preserved in the swap. Again OT ensures that every value is exactly decodable back to a full datum per individual in survey data. The 
interpreted visualisations remain unique. 

Recent state-space approaches to longitudinal survival analysis 
have similarly highlighted the importance of representing temporal 
evolution directly within the underlying system dynamics rather 
than solely through marginal hazard estimates~\cite{Cauchi2024}. 
Multistate modelling approaches formalise the transitions between 
absorbing and transient states in survival data~\cite{Putter2007}, 
providing a statistical parallel to the structural representation 
developed here.  OT is not limited to binary data. Categorical data can always be 
fit/unfit-encoded according to clearly defined purpose. Age of mothers is 
an example in the case of infant survival. Financial security or cost minimisation could equally serve 
as the fitness criterion in a public health context.  Where three-valued data is logical, OT may be reformulated, in principle. 
In such a case state space has $n!3^n$ states, increasing the difficulties 
of analysis in OT and in EHA, but not altering the qualitative properties 
found above.
 For multivariate binary longitudinal data of the kind 
analysed here, a natural workflow emerges: EHA identifies 
the statistically dominant variables and establishes which 
covariates survive adjustment for confounding and 
heterogeneity; OT, applied to that reduced variable set, 
exposes the structural organisation of trajectories and 
resolves differences that aggregate modelling cannot surface.


\section{Discussion of Results}
\label{sec:discussion} 

Table~\ref{tab:age_correspondence}  provides the quantitative
evidence linking the EHA age-hazard profile directly to the $S_3$
transition structure. Both representations agree that mortality risk
is highest in the first two years of life and declines monotonically
through ages 2--4; the EHA coefficient at age 1 is positive relative
to age 0 (consistent with a modest rise in mortality risk from year 0
to year 1 at the annual scale), while OT transition counts are highest
at age 0. This slight apparent discrepancy reflects that the EHA
coefficient is a log-odds ratio relative to age 0, not an absolute
count, whereas the OT column is an absolute transition count. The key
shared finding is the sharp decline from age 1 onward in both
representations. Mortality is concentrated in early childhood~\cite{You2015,UNIGME2024}. EHA and OT are in agreement on these results.   Further, both methods 
agree that the sibling variables (Table~\ref{tbl:vars}) 
do not contribute meaningfully to the mortality dynamics. 

The divergence between EHA and OT on maternal migration reflects data inhomogeneity rather than method error. EHA identifies maternal death as 
a robust predictor ($p<0.001$, Table~\ref{tab:f}) and 
finds maternal migration null ($p > 0.6$, Tables~\ref{tab:seffect}, \ref{tab:f}); OT identifies maternal migration 
as the organiser of 297 of the 340 observed sequential transitions into 
child-death states. Child death occurs in both MM states  (290 transitions 
when mother is present, 37 when absent), so the  conditional probability difference between the two configurations is small, and MM does not reach significance in EHA. Maternal death, by contrast, is so strongly absorbing 
and so rare (34 population-wide change-events) that it almost never appears 
as a preceding state in  a year-to-year transition; its role is terminal 
rather than  sequential, and EHA captures it through the absorbing geometry 
of the state space rather than through observed two-state transitions. 
These are structurally different causal roles, and the two methods surface 
them through different mechanisms -- both correctly, relative 
to their respective representations of the data.

A further reason for divergence between theories lies with length of 
time series per individual, which is known to place EHA in doubt ~\cite{Singer2003,BoxSteffensmeier2004,yamaguchi1991}. Uncertainty of the method is apparent from the improving 
$p$-values for age analysis of Tables~\ref{tab:nf} 
and~\ref{tab:f} as time series lengthen: the single child 
death observation in the first period drives unstable 
estimates at age 1. The change of sign across the CI is 
a strong signal of an unreliable result. Mother migration is seen in EHA 
to be unreliable as a covariate because the $p$-value is large and the 
CI again straddles the origin. 
 A further reason to question the EHA result on maternal migration 
is the extreme rarity of maternal death: only 34 population-wide 
change-events against 648 child deaths. 
The use of OT to identify objectively homogeneous clusters 
and microclusters is recommended for any multivariate binary 
longitudinal dataset, as a pre-analysis diagnostic before 
applying EHA. 
Variables with high $p$-values in EHA are not unreliable 
predictors per se -  they are unreliable only when the 
underlying data is inhomogeneous. OT makes this 
inhomogeneity directly visible, without statistical 
assumptions, through the congruence structure of observed 
transitions. 
  However, similar 
 patterns may a priori have different meaning and should not be aggregated. 

 The practical implication is that population-level
risk-factor identification, as EHA provides, is necessary but not
sufficient for targeting intervention. The reduced orbit system shows
that the 297 transitions preceding child death via maternal migration
reorganisation (Table~\ref{s3tab:trans}) constitute a structurally
distinct transition that aggregate hazard estimates do not surface.  
Maternal death appears as a changing variable in only 
2 of the 340 observed sequential transitions (Block III, 
Table~\ref{s3tab:trans}) - its structural role is terminal, 
not sequential. 
  Identifying these transition-specific subgroups -- not just the risk 
factors - is directly relevant for designing targeted health interventions. Recent work in data-driven modelling emphasises the importance of
retaining trajectory-level information in complex systems, particularly
where rare states drive system behaviour~\cite{Bzdok2022}.

\section{Conclusion}

 

Longitudinal binary data sets are rich in structure, and that 
structure is not accessible through standard statistical summaries alone. 
Plotting orbits reveals the structural complexity hidden in standard demographic summaries.  
  The full orbit space $S_{12}$ concentrates into 329 observed  states  
   reduced to 19 under  $\Phi$ in $S_3$. The projection recovers the dominant  mortality dynamics in 48 tractable states.  
   EHA is a powerful tool when data are homogeneous and time series are sufficiently long: low $p$-values and suitable confidence intervals identify associations and guide public health action.  
    The non-significance of maternal migration 
and the dominance of maternal death in EHA, taken together 
with the OT findings, reveal heterogeneity that frailty 
models do not fully capture. Summary statistics and EHA 
results -  whether variables associate or not -  are 
effective guides to both associations and the homogeneity 
requirements that underlie reliable aggregation.


  The reduced orbit system provides a state-space formulation consistent
with the principal findings of the EHA model.  The central contribution is the reduced orbit system defined 
by $(CS, MS, MM)$: 48 tractable states that preserve the 
dominant transition dynamics of the full $S_{12}$ space, expose 
structurally distinct high-risk clusters, and make the 
mechanisms of child death directly observable.  Maternal migration contributes to
trajectory reorganisation, and mortality risk remains concentrated in
early childhood. Additional structural features are revealed  - 
absorbing states, dominant transitions, and small high-risk
clusters  -  that cannot be recovered from regression-based analysis.  
  The reduced orbit system therefore emerges as a
structural concentration phenomenon rather than a visual
simplification.

The analysis here   demonstrates  that high-dimensional longitudinal processes can admit lower-dimensional representations that retain the essential dynamics.  Such reduction is facilitated by homogeneity in the observed transition structure.   The reduced orbit representation provides
a bridge between statistical modelling and objective state-space analysis 
of exact data,
offering a state-space-explicit alternative to sequence-analytic
approaches~\cite{Abbott1995,Billari2001,Aassve2007} for life-course
trajectory data: where optimal-matching methods summarise trajectories
by pairwise distance, OT preserves the full transition structure in an
enumerable state space.  The construction of $\Phi$ generalises to any
multivariate binary longitudinal dataset: variables are retained if
they (i) survive EHA-based covariate selection or (ii) demonstrably
participate in the transition structure of the orbit representation,
even where statistically null  -  as MM illustrates here. This makes
the approach applicable wherever multivariate binary longitudinal
records arise, particularly where the mechanisms of transition and
rare subgroup configurations are substantively important.

The substantive demographic findings presented here are specific  to the Agincourt HDSS and the 1998-2008 observation period,  predating the large-scale ART rollout. Their replication in 
other settings would require new data. 
Maternal death has been extensively studied at the Agincourt HDSS 
site, as reviewed in the Introduction.   A  limitation concerns missing data for maternal
migration status: person-periods with unrecorded migration were
excluded  from analysis rather than imputed, and the potential for
systematic missingness has not been formally assessed. 
 When observation gaps cause OT to collapse transitions, 
remaining states are still decodable back to the original 
data, but the effect of precedence is lost for those 
intervals.
If
out-migrating mothers are less likely to be observed, this
represents a conservative analytical choice whose directional effect
on the MM-related findings is unknown.  
 The analysis is specific to the
Agincourt HDSS, a rural South African population with a distinct
Mozambican refugee composition, and results may not generalise to other
demographic or geographic contexts. The binary coding required by OT
discards within-category distinctions  -  for example, between mother
age groups 25--29 and 30--34  -  and the sensitivity of the orbit
structure to this discretisation has not been systematically assessed.
OT also has no internal check for coding error: a misclassified
variable propagates through all downstream trajectory analysis, placing
a higher burden on data preparation than EHA does. The data span
1998--2008, predating the large-scale rollout of antiretroviral
treatment (ART) in South Africa, which accelerated substantially after
2004 and transformed the maternal and child mortality landscape in the
study region. The findings therefore describe a specific
epidemiological period rather than current conditions, and
generalisation to post-ART settings should be made with caution. At
the same time, the methodological contribution  -  the construction of
a reduced orbit system linking EHA and OT  -  is not dataset-specific
and applies to any multivariate binary longitudinal record regardless
of temporal period. The substantive findings are historical; the
structural method is general. 
 A limitation of OT is that, given the intricacy of a 
state space of $N = n!\cdot 2^n$ states, congruent 
transition sets may be overlooked without systematic 
numerical search procedures.

Several directions for future work emerge.  The structural 
mechanism by which EHA fails to detect sequentially changing
variables  - illustrated here by maternal migration -  
warrants formal investigation: the conditional probability 
argument in Section~\ref{sec:reduced} suggests a general 
result, applicable wherever a variable reorganises trajectories 
before rather than at the terminal event. OT transition counts 
on objectively homogeneous clusters yield empirical transition 
probability estimates that could serve as non-parametric 
specification checks on parametric hazard models; the 
connection to multistate survival models~\cite{Putter2007} 
provides one formal bridge. The congruence criterion for 
cluster aggregation provides an objective pre-analysis 
diagnostic for the homogeneity assumptions underlying EHA, 
and its formal development as a diagnostic tool is a direct 
extension of the present work. The construction of $\Phi$ 
generalises to any multivariate binary longitudinal dataset 
and warrants application in other domains where transition 
mechanisms and rare subgroup configurations are substantively 
important. Finally, point orbits -  children observed only 
once across the five-year window - contribute to state-space 
occupancy but carry no transition information; their entry and 
exit age distribution may affect the hazard profile in ways 
not yet formally examined.
 A natural inductive extension is to identify what state 
dominantly precedes the dominant death-preceding state, i.e.  extending the chain backwards, step by step, toward 
earlier precursors of child death.
 OT does not claim a unique path to child death  (other 
less-populated paths exist),  but it gives the dominant 
one, which is the probable path in the same sense as 
Markov analysis. 
 
\section*{Data availability}
The binarised data supporting this study are available at Figshare: 
\href{https://doi.org/10.6084/m9.figshare.34064565}{https://doi.org/10.6084/m9.figshare.34064565}. Requests to access the Agincourt HDSS individual-level data may be 
directed to the MRC/Wits Rural Public Health and Health Transitions 
Research Unit at \url{https://www.agincourt.co.za}. 


\section*{Author contributions}
W.L. conducted the Event History Analysis, produced the 
statistical models and associated tables, and contributed 
to manuscript writing. D.S. developed the Orbit Theory 
framework, contributed to methodological design, and 
supervised the analytical approach. M.V.V. conceptualised 
the study, led the Orbit Theory analysis and the 
comparative methodology, and coordinated manuscript 
preparation. All authors reviewed and approved the final 
manuscript.  

\section*{Funding}
The authors received no specific funding for this work.

\end{document}